\documentclass[a4paper,11pt]{article}
\pdfoutput=1 % if your are submitting a pdflatex (i.e. if you have
\usepackage{jcappub} % for details on the use of the package, please
\usepackage{amsmath}  
\usepackage{dsfont} 
\usepackage{comment}
\usepackage[T1]{fontenc} % if needed
\usepackage{enumerate}
\usepackage[color=orange]{todonotes}
\usepackage[normalem]{ulem}

 \usepackage[normalem]{ulem}
 \usepackage{hyperref}
\usepackage{academicons}
\usepackage{xcolor}
\usepackage{orcidlink}
\newcommand{\orcid}[1]{\href{https://orcid.org/#1}{\textcolor[HTML]{A6CE39}{\aiOrcid}}}

\title{\boldmath Preheating in  Einstein-Cartan Higgs Inflation: Oscillon formation}
\author[1]{Matteo Piani} %\orcidlink{0000-0002-2387-5948}~
\author[2]{and Javier Rubio} %\orcidlink{0000-0001-7545-1533}~
\affiliation[1]{Centro de Astrof\'{\i}sica e Gravita\c c\~ao  - CENTRA,
Departamento de F\'{\i}sica, Instituto Superior T\'ecnico - IST,
Universidade de Lisboa - UL, Av. Rovisco Pais 1, 1049-001 Lisboa, Portugal.}
\affiliation[2]{Departamento de F\'isica Te\'orica and Instituto de F\'isica de Part\'iculas y del Cosmos (IPARCOS-UCM), Universidad Complutense de Madrid, 28040 
Madrid, Spain} 
\emailAdd{matteo.piani@tecnico.ulisboa.pt}
\emailAdd{javier.rubio@ucm.es}
\abstract{We make use of classical lattice simulations in $3+1$ dimensions to study the preheating stage of Higgs Inflation in Einstein-Cartan gravity. Focusing for concreteness on a simplified scenario involving the seminal Nieh-Yan term, we demonstrate the formation of dense and spatially localized oscillon configurations constituting up to 70\% of the total energy density. The emergence of these meta-stable objects may lead to a prolonged period of matter domination, effectively modifying the post-inflationary history of the Universe as compared to the metric and Palatini counterparts. Notably, the creation of oscillons comes together with a significant gravitational wave signal, whose typical frequency lies, however, beyond the range accessible by existing and planned gravitational wave experiments. The impact of the Standard Model gauge bosons and fermions and the potential extension of our results to more general Einstein-Cartan settings is also discussed.}
\begin{document}
%%%%%%%%%%%%%%%%%%%%%%%%%%%%%%%%%%%%%%%%%%%%%%%%%%%%%%%%%%%%%%%%%%%%%%%%%%%%%%%%
\maketitle
\flushbottom

%%%%%%%%%%%%%%%%%%%%%%%%%%%%%%%%%%%%%%%%%%%%%%%%%%%%%%%%%%%%%%%%%%%%%%%%%%%%%%%%
\section{Introduction} \label{sec:intro}
%%%%%%%%%%%%%%%%%%%%%%%%%%%%%%%%%%%%%%%%%%%%%%%%%%%%%%%%%%%%%%%%%%%%%%%%%%%%%%%%

The constant improvement in the precision of measurements of the Cosmic Microwave Background (CMB)~\cite{Planck:2018jri,BICEP:2021xfz} has led to the establishment of the inflationary paradigm~\cite{Guth:1980zm,Linde:1981mu,Mukhanov:1981xt} as the best candidate to tackle the conceptual problems of the hot Big Bang while generating the primordial density perturbations seeding structure formation. Despite this success, the nature of the inflaton field remains unknown and its role could be played by any candidate able to mimic a scalar field in the slow-roll regime. Indeed, while the Planck and BICEP2 collaborations have set strong constraints on the scale dependence of the density power spectrum and the tensor-to-scalar ratio encoding the amount of primordial gravitational waves \cite{BICEP:2021xfz}, large classes of inflationary models remain still compatible with observations \cite{Kallosh:2019hzo}. On top of that, knowing the shape of the inflationary potential is a priori not sufficient to determine the whole post-inflationary evolution. In particular, any inflationary model must come together with a graceful exit mechanism able to transfer the energy stored in the homogeneous inflaton condensate to the Standard Model (SM) degrees of freedom, leading with it to the onset of the usual hot Big Bang theory \cite{Bassett:2005xm,Allahverdi:2010xz, Amin:2014eta}. This poses significant limitations for many particle physics scenarios in which the inflaton-matter couplings are not experimentally known.

Among the many possible particle physics embeddings of the early Universe accelerated expansion, Higgs Inflation (HI)~\cite{Bezrukov:2007ep,Rubio:2018ogq} stands out as an ideal candidate for studying the post-inflationary dynamics. In fact, assuming no Beyond SM physics between the electroweak and the Planck scale, it is a priori possible to compute the running of the experimentally known SM couplings, allowing, in principle, for a comprehensive study of the preheating stage. This scenario has been studied in different formulations of gravity, ranging from the standard metric case~\cite{Bezrukov:2007ep,Barbon:2009ya,Bezrukov:2009db,Burgess:2010zq,Bezrukov:2010jz,Giudice:2010ka,Bezrukov:2014bra,Hamada:2014iga,Hamada:2014wna,George:2015nza,Fumagalli:2016lls,Bezrukov:2017dyv} to the Palatini~\cite{Bauer:2008zj,Bauer:2010jg,Rasanen:2017ivk,Enckell:2018kkc,Rasanen:2018fom,Gialamas:2019nly,Shaposhnikov:2020fdv,Annala:2021zdt}, teleparallel~\cite{Raatikainen:2019qey}, Einstein-Cartan~\cite{Langvik:2020nrs,Shaposhnikov:2020gts,Shaposhnikov:2020aen,Karananas:2021zkl} and metric-affine formulations~\cite{Azri:2017uor,Rigouzzo:2022yan,Gialamas:2022xtt}, including also several scale-invariant extensions \cite{Shaposhnikov:2008xb,Garcia-Bellido:2011kqb,Garcia-Bellido:2012npk,Bezrukov:2012hx,Rubio:2014wta,Trashorras:2016azl,Casas:2017wjh,Tokareva:2017nng,Casas:2018fum,Shaposhnikov:2018jag,Herrero-Valea:2019hde,Karananas:2020qkp,Rubio:2020zht,Shaposhnikov:2020frq,Karananas:2021gco,Gialamas:2021enw,Piani:2022gon,Belokon:2022pqf}. Conversely, the post-inflationary evolution has been studied both analytically and numerically only for the metric~\cite{Bezrukov:2008ut,Garcia-Bellido:2008ycs,Bezrukov:2014ipa,Repond:2016sol,DeCross:2015uza,DeCross:2016fdz,DeCross:2016cbs,Ema:2016dny,Sfakianakis:2018lzf} and Palatini~\cite{Rubio:2019ypq,Dux:2022kuk} cases.

When matter is disregarded it is always possible to map all these different formulations back to the metric one. However, this ceases to be true in the presence of fermions and non-minimal couplings with scalar fields (see~\cite{Rigouzzo:2022yan,Rigouzzo:2023sbb} for a detailed explanation). Such feature becomes explicit when turning to the Einstein-frame, where the Lagrangian reduces to the one of metric gravity, plus a modified potential and a non-canonical kinetic term for the Higgs field. Although one could argue that such scenario could be also attained by directly modifying the minimally coupled Standard Model Higgs Lagrangian in the metric formulation, the choice of operators in that case would remain arbitrary. On the contrary, the shape of the theory defining functions in the framework under consideration is completely determined by gravitational interactions, naturally providing a set of selection rules for the higher-order operators in the Einstein-frame. 
 In this context, the specific formulation one uses provides a selection criterium which leads to different outcomes, therefore breaking the degeneracy between all the possible gravitational theories.

In this work, we make use of fully-fledged numerical lattice simulations in $3+1$ dimensions to study the preheating stage of HI in Einstein-Cartan~(EC) gravity. Restricting ourselves to the radial degree of freedom of the Higgs field, we find that, contrary to what happens in the metric and Palatini formulations of the theory, this scenario allows for the formation of dense and spatially-localized oscillon configurations constituting up to $70\%$ of the total energy density for suitable model parameters. In spite of lacking a conserved charge associated with them, these pseudo-solitonic objects can be rather long-lived, leading to a prolonged matter-domination era that effectively alters the minimum duration of the inflationary phase needed to solve the usual hot Big Bang problems. Additionally, we find that such structures can source a sizeable gravitational wave signal, thus providing an alternative observational channel for this appealing inflationary model.  The emergence of oscillons is expected to be robust against the inclusion of gauge fields, since the accumulation of these species turns out to be severely suppressed by their perturbative decay into fermions for a large number of oscillations, significantly exceeding the oscillon formation time. 

The paper is structured as follows. In Section~\ref{sec:1} we review the embedding of HI in the EC gravity, highlighting on the differences with the metric and Palatini formulations. Section~\ref{sec:2} is devoted to the study of the preheating stage in the scalar sector of theory, starting from a simple linear analysis to subsequently perform a fully non-linear characterization of oscillon formation via 3+1 classical lattice simulations. Aspects such as the abundance and energy distribution of oscillons or the associated production of gravitational waves at formation are addressed in detail, comparing the latter with the sensitivity and frequency range of current and future GWs experiments. The potential impact of the SM gauge boson and fermions is considered in Section~\ref{sec:4}.  Finally, our conclusions are presented in Section~\ref{sec:conclusions}, where we argue on possible extensions and improvements of our treatment. 

%%%%%%%%%%%%%%%%%%%%%%%%%%%%%%%%%%%%%%%%%%%%%%%%%%%%%%%%%%%%%%%%%%%%%%%%%%%%%%%%
\section{Higgs Inflation in Einstein-Cartan Gravity}\label{sec:1}
%%%%%%%%%%%%%%%%%%%%%%%%%%%%%%%%%%%%%%%%%%%%%%%%%%%%%%%%%%%%%%%%%%%%%%%%%%%%%%%%

The EC formulation of gravity assumes the tetrads $e^A_\mu$ and spin connection $\omega_\mu^{AB}$ ${(\omega _ \mu ^ {AB} = - \omega _ \mu ^ {BA}})$ as fundamental gravitational variables \cite{Utiyama:1956sy,Kibble:1961ba,Hehl:1976kj,Shapiro:2001rz}, with the Greek letters related to the usual spacetime coordinate basis and the Latin ones to an orthonormal non-coordinate basis displaying covariance under local Lorentz transformations. The associated metrics and connections entering the covariant derivatives in the two bases,
\begin{equation}
D_\mu V^\alpha = \partial_\mu V^\alpha + \Gamma^\alpha_{\sigma\mu} V^\sigma\,, \hspace{10mm}
D_\mu V^A = \partial_\mu V^A + \omega_\mu^{AB} V_B \,,
\end{equation}
are related through 
\begin{equation}
g_{\mu\nu}=e^A_\mu e^B_\nu \eta_{AB}\,,
\hspace{10mm} 
\Gamma^\mu_{\;\;\nu\rho}= e^\mu_A\left( \partial_\rho e^A_\nu+\omega^A_{\rho B} 
e^B_\nu \right) \,,
\end{equation}  
with the second expression following directly from the condition $D_\mu e_\nu ^ A=0$, without any symmetry assumption on the lower indices. Beyond the curvature tensor, this theory accounts for a non-vanishing torsion contribution 
\begin{equation}\label{eq:torsion}
T^\mu_{\;\;\nu\rho}=\Gamma^\mu_{\;\;\nu\rho}-\Gamma^\mu_{\;\;\rho\nu}\,,
\end{equation}
obtained by acting with the commutator of two covariant derivatives on a given scalar (or a vector, upon subtracting curvature contributions).

As compared to General Relativity, this construction offers some conceptual advantages. Firstly, EC gravity  can be viewed as the gauge theory of the Poincaré group, placing gravitational interactions on equal footing with other fundamental forces in Nature \cite{Utiyama:1956sy,Kibble:1961ba}. Secondly, the independent treatment of the metric and the connection in EC gravity avoids the necessity of introducing specific boundary terms in order to obtain the Einstein equations of motion \cite{Ferraris1982}. Finally, the spin connection can be easily coupled to fermions, making the EC formulation particularly useful in describing the behaviour of the SM in the presence of gravity.

In the context of scalar-tensor theories, the existence of torsion opens up the possibility of including additional gravitational operators beyond those usually considered in metric HI, namely the usual SM Higgs sector and the Einstein-Hilbert term modified by the addition of a non-minimal coupling to gravity,~\footnote{We adopt here a mostly plus signature and work in Planckian units $M_P^2\equiv 8\pi G=1$. }
\begin{equation}\label{S-EH}
S_{\rm SM + EH}  =\int d^4x\sqrt{-g} \left[ \frac{1+\xi h^2}{2} g^{\mu\nu}R_{\mu\nu}(\Gamma)-\frac{1}{2}g^{\mu\nu}\partial_\mu h \partial_\nu h-\frac{\lambda}{4}(h^2-v^2)^2\right]\,,
\end{equation} 
with $h$ the Higgs field in the unitary gauge, $\xi$ a dimensionless coupling constant to be determined from observations, $\lambda$ the Higgs self-coupling, and $v\simeq 250$ GeV its vacuum expectation value. In the context of EC gravity, building the most general theory containing only a massless spin-2 gravitational degree of freedom requires the introduction of a plethora of operators. As shown explicitly in Ref.~\cite{Karananas:2021zkl} (see also Refs.~\cite{Langvik:2020nrs,Shaposhnikov:2020gts} for specific choices), 
the most general graviscalar sector beyond \eqref{S-EH} including at most quadratic operators in the derivatives of all fields takes the form 
\begin{equation}
\begin{aligned}
\label{eq:action_torsion}
S_{\rm T} &= \int d^4 x \,\sqrt{-g}\Bigg[ v^\mu \partial_\mu Z^v + a^\mu \partial_\mu Z^a \\
&+ \frac{1}{2}\Big(G_{vv} v_\mu v^\mu  + 2G_{va}v_\mu a^\mu + G_{aa}	a_\mu a^\mu 
 +G_{\tau\tau}\tau_{\alpha\beta\gamma} \tau^{\alpha\beta\gamma}+ \tilde{G}_{\tau\tau} \epsilon^{\mu \nu \rho \sigma} \tau_{\lambda\mu\nu} \tau^\lambda_{~\rho\sigma}\Big) \Bigg]\,,
\end{aligned}
\end{equation}
with 
\begin{equation}
Z^{v/a}=\zeta^{v/a}_h h^2 \,,  \hspace{15mm} 
G_{ij}= c_{ij} \left(1+\xi_{ij}h^2\right) \ ,
\end{equation} 
no summation repeated $i,j$ indices and $\xi^{v/a}_h$, $c_{ij}$ and $\xi_{ij}$ constants. Here
\begin{equation}
v_\mu = T^\nu_{~\mu\nu} \,, \hspace{10mm} 
a_\mu = \epsilon_{\mu\nu\rho\sigma}T^{\nu\rho\sigma} \,,
\hspace{10mm}
\tau_{\mu\nu\rho} =\frac 2 3 \left( T_{\mu\nu\rho} -v_{[\nu} g_{\rho]\mu} - T_{[\nu\rho]\mu} \right) \ ,
\end{equation}
stand respectively for the vector, pseudo-vector and reduced tensor irreducible components of the torsion tensor \eqref{eq:torsion},
\begin{equation}
\label{eq:tors_irreps}
T_{\mu\nu\rho} = e_{\mu A} T^A_{\nu\rho}= \frac23 v_{[\nu}g_{\rho]\mu} - \frac16 a^\sigma \epsilon_{\mu\nu\rho\sigma} +\tau_{\mu\nu\rho} \,,
\end{equation}
with the square brackets standing for anti-symmetrization in the corresponding indices.  

In order to get a more intuitive picture of the inflationary and post-inflationary dynamics in this rather complicated theory, it is convenient to move to the Einstein-frame, where the Higgs field is minimally coupled to gravity and no explicit torsion operator is present. This can be achieved in two steps. First, we perform a Weyl rescaling of the metric with $g_{\mu\nu}\rightarrow (1+\xi h^2)^{-1}g_{\mu\nu}$. Second, we solve the equation of motion for the anti-symmetric part of the connection, \textit{i.e.} the one encoding torsion, and plug back the solution into the action. The explicit computation has already been developed in Refs.~\cite{Karananas:2021zkl, Langvik:2020nrs,Shaposhnikov:2020frq}, to which we refer to the interested reader for details.  The final action $S=S_{\rm SM+EH}+S_{\rm T}$ takes the compact form
\begin{equation}\label{eq:EF-Action} 
S =\int d^4 x \sqrt{-g} \Bigg[ \frac{R}{2} - \frac{1}{2} K(h)(\partial h)^2  - V(h)\Bigg] \,,
\end{equation}
with
\begin{equation}\label{eq:NY-kinetic}
K (h)=\frac{1+ c \, h^2}{(1+\xi h^2)^2}\,,\hspace{15mm} V(h)=\frac{\lambda}{4}\frac{h^4}{(1+\xi h^2)^2}\,,
\end{equation}
and
\begin{equation}\label{eq:cgeneric}
c(h)= \xi + 6\xi^2 +4 f(h) \frac{G_{aa}(\zeta^{v}_h)^{2}+G_{vv}({\zeta^{a}_h})^{2}-G_{va}{\zeta^{v}_h} {\zeta^{a}_h}}{G_{vv}G_{aa}-G_{va}^2}\;.
\end{equation}
Here we have explicitly neglected the Higgs vacuum expectation value, since this will not be relevant for our analysis of inflation and preheating. 
For the purposes of this paper, we will mainly be interested in EC scenarios leading to field-independent $c$ values. For illustration purposes only, in what follows we will consider a simple scenario with  
\begin{equation}
\begin{aligned}
& c_{vv} = -\frac{2}{3}\,, \hspace{10mm} c_{va} =0\,, \hspace{10mm}   c_{aa}=\frac{1}{24}\,, \\ 
& \xi_{vv}=\xi_{aa}=-\zeta^{v}_h=\xi\,, 
\hspace{10mm} \xi_{va}=0\,, \hspace{10mm}
\zeta^{a}_h=\frac{1}{4}\xi_\eta \;,
\end{aligned}
\end{equation}
leading effectively to a parameter
\begin{equation}
c=\xi + 6\xi^2_\eta\,. 
\end{equation}
In this limit, the general equation \eqref{eq:action_torsion} reduces to a simple interaction term involving the seminal Nieh-Yan topological invariant~\cite{Nieh:1981ww,Nieh:2008btw}
\begin{equation}
    \label{eq:Nieh-Yan}
    S_{\rm T} =- \frac{1}{4}\int d^4x\:\xi_\eta h^2\partial_{\mu}\left(\sqrt{-g}\epsilon^{\mu\nu\rho\sigma}T_{\nu\rho\sigma}\right)~,
\end{equation} 
with $\epsilon^{\mu\nu\rho\sigma}$ the totally anti-symmetric tensor ($\epsilon_{0123}=1$) and $\xi_\eta$ a non-minimal coupling to be determined from observations. As shown in Refs.~\cite{Langvik:2020nrs,Shaposhnikov:2020gts,Piani:2022gon}, this specific choice provides a smooth parametric interpolation between the metric ($\xi_\eta=\xi$) and Palatini formulations of HI ($\xi_\eta=0$), a property that, as we will show in Section \ref{sec:2}, will turn out to be critical for oscillon formation.  Note, however, that most of our results can be easily extended to more general settings within the EC multiverse \eqref{eq:action_torsion}, provided that they effectively lead to a constant value of $c$ in the field range of interest. 

For constant $c$, the non-canonical kinetic term in Eq.~\eqref{eq:EF-Action} can be made canonical by performing a field redefinition
\begin{equation}\label{eq:inf-canonical-field}
 \frac{d\chi}{dh}=\sqrt{K(h)}\,.
\end{equation}
Solving this differential equation and inverting the result in the small ($c\,h^2 \ll 1$) and large field ($c\,h^2 \gg 1$) regimes, we obtain the effective Einstein-frame potential
\begin{equation}\label{eq:potential}
V \simeq \begin{cases} 
\dfrac{\lambda}{4}\chi^4\,, &  \hspace{5mm} \textrm{for} \hspace{5mm} \chi<\chi_c~\,,  \\
\dfrac{\lambda}{4 \xi^2}\left[1-\exp\left(- \dfrac{2\xi|\chi|}{\sqrt{c}}\right) \right]^{2}\,,& \hspace{5mm} \textrm{for} \hspace{5mm} \chi \gg\chi_c\,,
   \end{cases}
\end{equation}
with $\chi_c=1/\sqrt{c}$ a crossover scale depending only on the parameter $c$. This expression flattens out exponentially at large field values, allowing for inflation to take place with the usual slow-roll initial conditions. In this regime, our scenario coincides essentially with that of $\alpha$-attractor E-models \cite{Kallosh:2021mnu}, upon the formal replacement $c\to 6 \xi^2\alpha$, but in this case with a discrete $\mathds{Z}_2$ symmetry. Although such a reflection invariance does not affect at all the inflationary dynamics (other than allowing, of course, for inflation with negative and positive field values), it will have important consequences for the preheating phase, as we will show explicitly in Section \ref{sec:2}.

\begin{figure}
\begin{center}
 \includegraphics[width=0.8\textwidth]{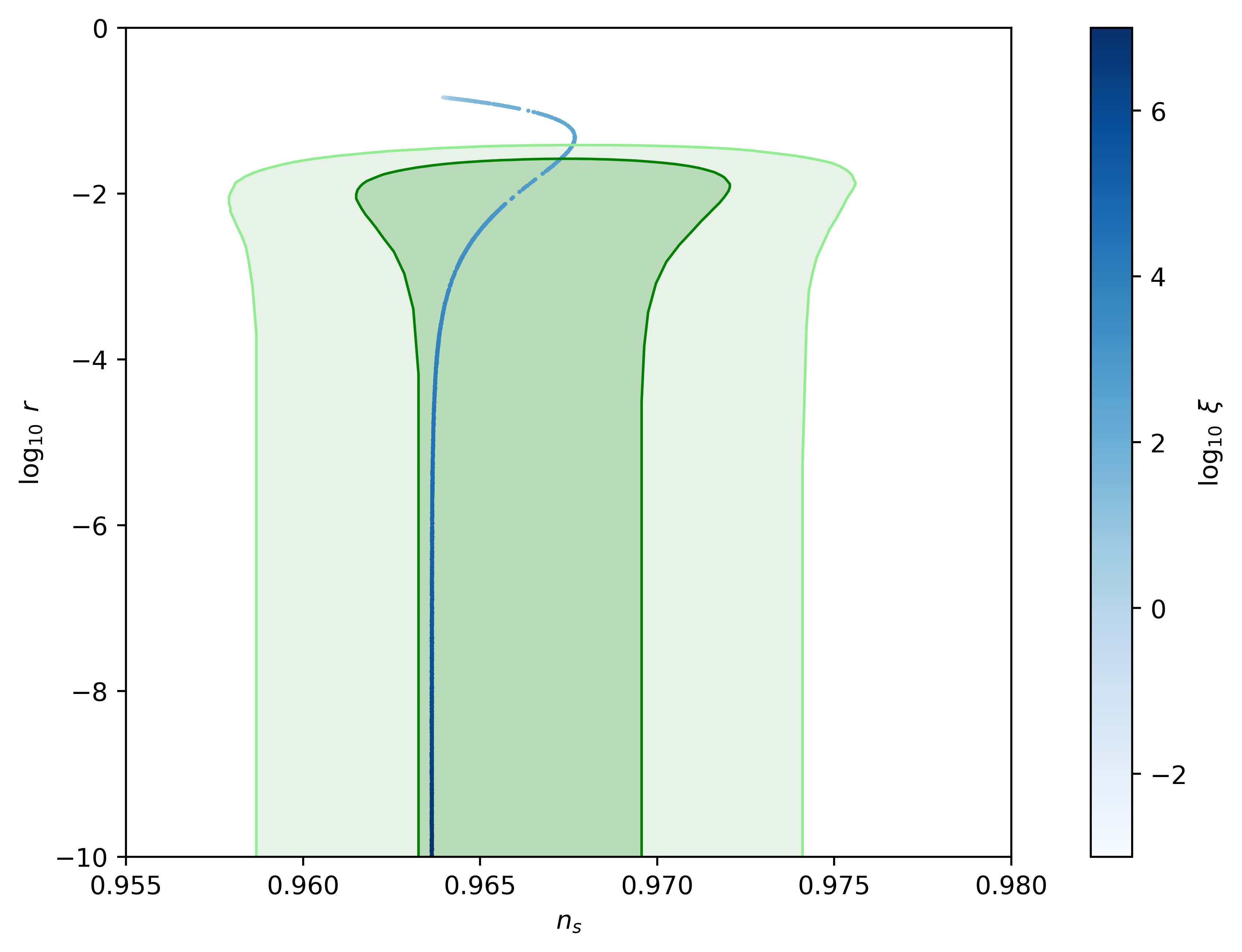} 
\caption{Numerical results (blue curve) in the $n_s-r$ plane for the action~\eqref{eq:EF-Action} with different values of the non-minimal coupling $\xi$ (following the procedure in~\cite{Shaposhnikov:2020gts,Piani:2022gon}), together with the recently updated constraints (green shaded regions) from the BICEP/Keck collaboration \cite{BICEPKeck:2022mhb}, at 68\% and 95\% C.L.} 
\label{fig:ns-r}  
 \end{center}
\end{figure}

Knowing the potential in canonical field variables, we can proceed now with the standard inflationary analysis, computing explicitly the slow-roll parameters,
\begin{equation}
    \label{eq:slow-roll-parameters}
    \epsilon\equiv \dfrac{1}{2}\left(\dfrac{V_{,\chi}}{V}\right)^2\simeq\dfrac{8 \xi ^2}{c}\exp\left(- \dfrac{4\xi|\chi|}{\sqrt{c}}\right)~,\hspace{10mm} 
    \eta\equiv \dfrac{V_{,\chi\chi}}{V}\simeq -\frac{8 \xi ^2 }{c} \exp\left(- \dfrac{2\xi|\chi|}{\sqrt{c}}\right)\,,
\end{equation}
and the relation between the inflaton field value and the number of $e$-folds of inflation, 
\begin{equation}
    \label{eq:number-of-efolds}
   {\cal N}(\chi)=\int_{\chi_{\text{end}}}^\chi \frac{d \chi'}{\sqrt{2\epsilon(\chi')}}  \simeq \frac{c}{8 \xi ^2}  \exp\left(\dfrac{2\xi|\chi|}{\sqrt{c}}\right) \hspace{5mm}\longrightarrow \hspace{5mm} \vert\chi(\mathcal{N})\vert =\frac{\sqrt{c}}{2 \xi }\log \left(\frac{8 \xi ^2 \mathcal{N}}{c}\right)~,
\end{equation}
with 
\begin{equation}  \label{eq:end-of-inflation}
    \chi_{\text{end}}=\frac{\sqrt{c}}{2 \xi } \log \left(1+\frac{2 \sqrt{2} \xi }{\sqrt{c}}\right)
\end{equation}
the field value at which inflation ends [$\epsilon(\chi_{\text{end}})\equiv 1$]. This allows us to determine the amplitude of the primordial spectrum of density perturbations $A_s=V/(24\pi^2 M_P^4 V)$, its tilt $n_s=1+2\eta-6\epsilon$ and the tensor-to-scalar ratio $r=16 \epsilon$ at the time at which the pivot scale $k_*=0.05\,{\rm Mpc}^{-1}$ exits the horizon (see figure \ref{fig:ns-r}),
\begin{equation}
    \label{eq:inflationary-parameters}
A_s=\frac{\lambda \mathcal{N}_*^2}{12 \pi^2 c}\,,\hspace{12mm}    n_s=1-\frac{2}{\mathcal{N}_*}~,\hspace{12mm} r=\frac{2 c }{\xi^2 \mathcal{N}_*^2}\,.
\end{equation}
The precise value of the number of $e$-folds ${\cal N}_*$ to be inserted in these expressions depends on the whole post-inflationary evolution, and in particular, on how long it takes after inflation to enter the phase of radiation domination needed for successful Big Bang Nucleosynthesis.  Note also that the observational constraint on the CMB-normalization, $A_s=2.1 \cdot 10^9$ \cite{Planck:2018jri}, 
restricts only the ratio of couplings determining the amplitude of the inflationary potential, namely
\begin{equation}\label{eq:c_over_lambda}
\frac{c}{\lambda}= \frac{2}{5}\cdot 10^{7} \, \mathcal{N}_*^2 \,.
\end{equation}
Therefore, as far as inflation is concerned, a change in the Higgs self-coupling $\lambda$ can be always compensated by a change of the parameter $c$, meaning that a strict relation between cosmological observables and SM parameters can only be established once the running of $\lambda$ is known. In the context of Einstein-Cartan gravity, this is expected to depend not only on the usual SM couplings but also on the actual strength of five- and six- dimensional Higgs-fermion and fermion-fermion interactions appearing when torsion is integrated out \cite{Kibble:1961ba}, potentially extended with additional non-minimal couplings \cite{Freidel:2005sn,Alexandrov:2008iy,Diakonov:2011fs,Magueijo:2012ug,Shaposhnikov:2020aen,Karananas:2021zkl}. In the lack of a proper computation of the SM running in this context, we will consider in what follows a fiducial value $\lambda=10^{-3}$. This choice is commensurate with the value obtained using the pure SM running for the top quark \textit{pole} masses in Refs.~\cite{ATLAS:2019guf,CMS:2019esx}, while coinciding also with the favoured one in the Palatini formulation of HI \cite{Shaposhnikov:2020fdv}.~\footnote{For a discussion of HI for negative values of $\lambda$, we refer the reader to Refs.~\cite{Bezrukov:2014ipa,Rubio:2015zia}.}

A simple inspection of the inflationary observables \eqref{eq:inflationary-parameters} reveals several interesting regimes.  For $\xi=\xi_\eta$, one recovers the metric HI predictions, where both the spectral tilt and the tensor-to-scalar ratio are almost insensitive to the specific value of the non-minimal couplings to gravity. On the contrary, for $\xi \gtrsim \xi_\eta^2$, we rather have $c\simeq \xi$, recovering effectively the predictions of the Palatini HI scenario, and in particular the strong $1/\xi$ suppression in the tensor-to-scalar ratio. The intermediate regime, $\xi_\eta < \xi <\xi_\eta^2$, leads to a tensor-to-scalar ratio proportional to $\xi_\eta^2/\xi^2$, with $\xi_\eta$ fixed to a value $\xi_\eta\simeq \sqrt{2\lambda/3}\cdot 10^3 \, \mathcal{N}_*$ by Eq.~\eqref{eq:c_over_lambda} and $\xi$ almost independent of the CMB normalization. Since preheating has already been extensively studied both in the metric~\cite{Bezrukov:2008ut,Garcia-Bellido:2008ycs,Bezrukov:2014ipa,Repond:2016sol,DeCross:2015uza,DeCross:2016fdz,DeCross:2016cbs,Ema:2016dny,Sfakianakis:2018lzf} and Palatini~\cite{Rubio:2019ypq,Dux:2022kuk} regimes, we will focus in what follows on the uncharted range $\xi_\eta < \xi <\xi_\eta^2$. 

%%%%%%%%%%%%%%%%%%%%%%%%%%%%%%%%%%%%%%%%%%%%%%%%%%%%%%%%%%%%%%%%%%%%%%%%%%%%%%%%
\section{Preheating in the scalar sector}\label{sec:2}
%%%%%%%%%%%%%%%%%%%%%%%%%%%%%%%%%%%%%%%%%%%%%%%%%%%%%%%%%%%%%%%%%%%%%%%%%%%%%%%%

Immediately after the end of inflation, the vast majority of the energy density of the Universe is stored in the zero mode of the Higgs field, which will start oscillating around the minimum of the potential \eqref{eq:potential}, leading with it to periodically time-varying masses for all the SM particles coupled to it and potentially inducing particle creation. The direct creation of SM fermions out of the Higgs condensate is, however, severely restricted by Pauli blocking effects \cite{Greene:2000ew,Peloso:2000hy}, leaving the production of Higgs  excitations and electroweak gauge bosons as main depletion channels. 

As a first step to study the complicated preheating dynamics of EC HI, let us concentrate on the pure Higgs sector, neglecting momentarily its coupling to the SM gauge bosons and fermions. In this regime, the evolution of the inflaton field in a Friedmann-Lema\^itre-Robertson-Walker background obeys a single-field Klein-Gordon equation
\begin{equation}
\label{eq:KG-inflaton}
\ddot{\chi}+3 H\dot{\chi}-a^{-2}\nabla^2 \chi+V_{,\chi}=0~, 
\end{equation}
with $a$ the scale factor, the dots denoting derivatives with respect to the cosmic time $t$ and the Hubble rate $H=\dot a/a$ evolving according to the Friedmann equations 
\begin{equation}
\label{eq:Friedmanneq}
	3 H^2 =\frac{1}{2}\dot{\chi}^2+ \frac{1}{2a^2}(\nabla\chi)^2  + V(\chi) \, , \hspace{10mm} \quad \dot{H} = -\frac{1}{2} \dot{\chi}^2-\frac{1}{6a^2} (\nabla\chi)^2 \,.
\end{equation}
The initial stages of preheating can be understood by splitting the inflaton field into a homogeneous component $\bar \chi(t)$ satisfying the zero-gradient limit of Eqs.~\eqref{eq:KG-inflaton} and \eqref{eq:Friedmanneq} and a small perturbation $\delta\chi({\bf x},t)$. The combination of the background equations of motion leads to the rather enlightening expression \cite{Rubio:2019ypq}
\begin{equation}\label{Hrate}
 \frac{dH}{d\bar h} = -\frac{\dot{\bar\chi}^2}{2\dot{\bar h}} = -\frac{1}{2}\sqrt{6H^2  - 2V[\bar \chi(\bar h)]}\frac{d\bar \chi}{d\bar h} \, ,
\end{equation}
with $V[\bar \chi(\bar h)]$ the $h$-dependent potential in Eq.~\eqref{eq:NY-kinetic} and the field redefinition $d\bar \chi/d\bar h$ completely encoding the specific dependence on the EC formulation of gravity under consideration, cf.~Eq.~\eqref{eq:inf-canonical-field}. A simple inspection of the associated kinetic function \eqref{eq:NY-kinetic} at the intermediate field values $\bar h_c \ll \bar h \ll \bar h_{end}$ reveals different behaviours for different values of $c$. In particular, for $c\rightarrow \xi+6 \xi^2$, corresponding to the metric HI limit $\xi_\eta=\xi$, the kinetic function remains parametrically large,  $d\bar \chi/d\bar h\approx \sqrt{6}\xi \bar h/M_P \gg 1$, leading to a quick damping of the oscillations as $\bar h$ approaches zero. On the other hand, for $c \rightarrow \xi$, corresponding to the Palatini HI limit $\xi_\eta=0$, we have approximately $d\bar \chi/d\bar h\approx 1$, making the damping of the oscillations significantly less efficient and allowing for the inflaton field to return closer to the inflationary plateau for a given number of oscillations. These recursive field incursions have a dramatic effect on the production of Higgs excitations, as becomes evident when considering the mode equations for the linear perturbations $\delta \chi$ in Fourier space,
\begin{equation}
    \label{eq:perturbation growth}
  \ddot{\delta \chi_{\textbf{k}}}+3 H\dot{\delta \chi_{\textbf{k}}} +\left(\textbf{k}^2/a^2 +V_{,\chi\chi}\right)\delta \chi_{\textbf{k}}=0\,,
\end{equation}
with $\bf k$ the associated wave-vector. In particular, the effective square frequency of this damped harmonic oscillator becomes non-positive definite whenever the Higgs field accelerates towards the minimum of the potential from beyond the inflection point $\chi_i =\sqrt{c}/(2 \xi)\ln 2$. This translates into a tachyonic amplification of long-wave fluctuations ($\textbf{k}^2/a^2 <\vert V_{,\chi\chi}\vert$) that can significantly enhance the efficiency of other Bose stimulation effects. 
The maximum momentum that will grow due to tachyonic instability is given by the minimum of the second derivative $V_{,\chi\chi}$, which turns out to be located at a parametrically fixed value
\begin{equation}
    \chi_{\rm min}=\frac{\sqrt{c} \log(2)}{\xi}\,,
\end{equation}
such that 
\begin{equation}\label{eq:max-momentum}
\frac{|\mathbf{k}_{\rm max}|}{a}=\sqrt{|V_{,\chi,\chi}(\chi_{\rm min})|}=2^{-3/2} M \simeq 0.35 M\,,
\end{equation}
with 
\begin{equation}\label{eq:rescaling}
    M^2\equiv V_{,\chi\chi}\vert_{\chi_c\rightarrow0}=\frac{2 \lambda }{c}
\end{equation} 
the oscillation frequency of the condensate for $\chi_c\rightarrow 0$. 

\begin{figure}
\begin{center}
 \includegraphics[width=0.8\textwidth]{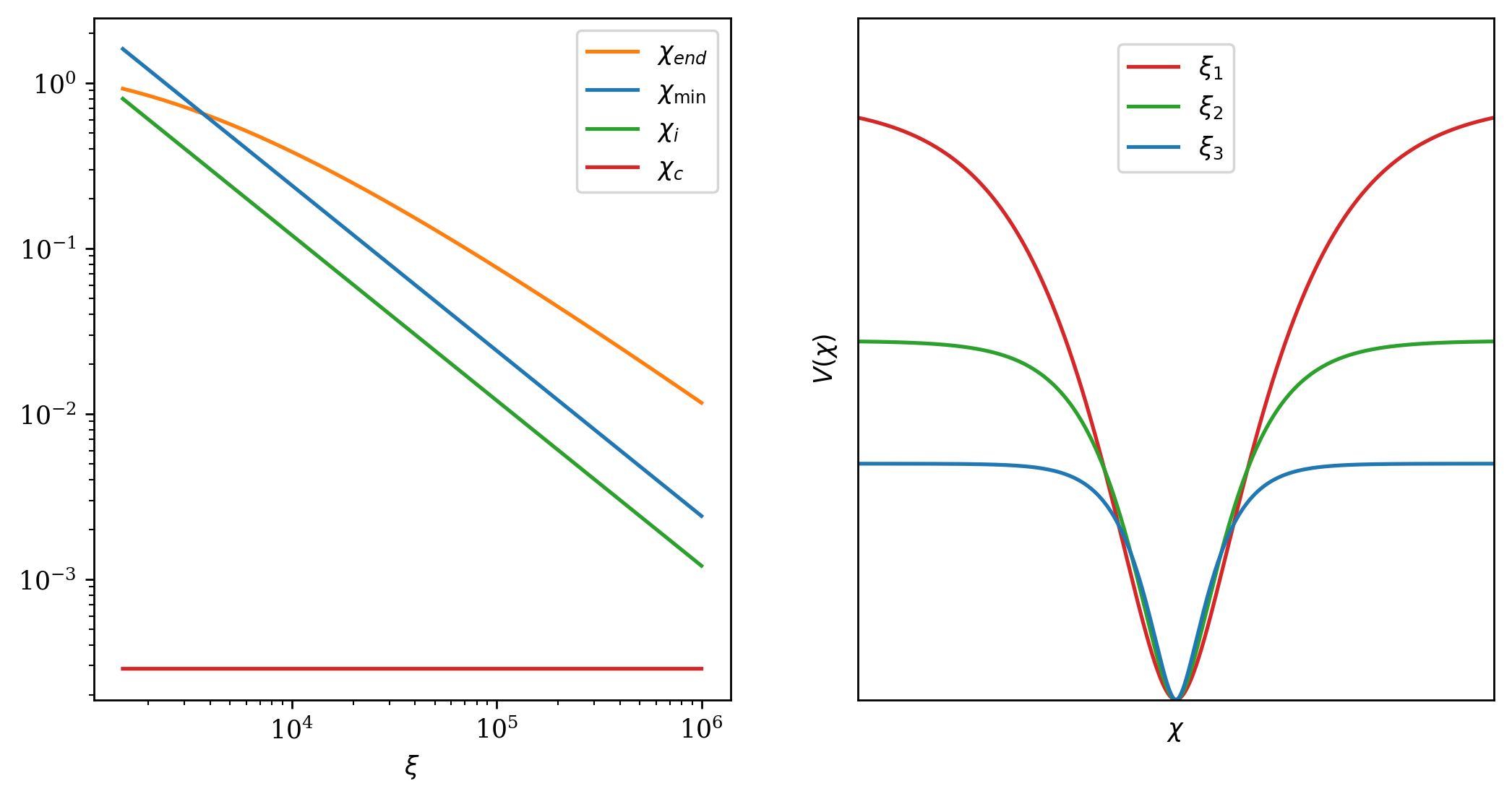} 
  \caption{(Left) Field values for the end of inflation $\chi_{\rm end}$ (orange), the location of the inflection point $\chi_i$ (green) and the field value associated to the minimum of $V_{,\chi\chi}$ (or the maximum momentum growing due to tachyonic instability) as a function of the non-minimal coupling $\xi$. The red line indicates the critical field value $\chi_c$ below which the potential becomes quartic. As expected, in the metric limit $\xi\simeq 1500$ inflation ends too close to the inflection point, which combined with a higher Hubble friction prevents the field to re-enter the tachyonic region of the potential in successive oscillations. (Right) Schematic form of the Einstein-frame HI potential for different values of the non-minimal coupling $\xi$, with $\xi_1<\xi_2<\xi_3$.} \label{fig:xi-dependence}  
 \end{center}
\end{figure}

If the resulting fluctuations are not efficiently diluted by the expansion of the Universe, they will eventually backreact on the homogeneous dynamics, leading to the fragmentation of the condensate into a highly inhomogeneous state with sizable overdensities. These overdensities can potentially collapse under their own self-interactions, leading to the formation of periodically-varying pseudo-solitonic objects or oscillons \cite{Bogolyubsky:1976yu,Gleiser:1993pt,Copeland:1995fq,Amin:2010dc,Amin:2011hj,Lozanov:2017hjm,Lozanov:2019ylm}. Indeed, such localized quasi-spherical configurations are known to appear in many inflationary models involving a shallower-than-quadratic potential \cite{Lozanov:2017hjm,Hasegawa:2017iay,Mahbub:2023faw,Sang:2019ndv,Sang:2020kpd},\footnote{This feature allows for the partial cancellation of the dispersion terms in Eq.~\eqref{eq:KG-inflaton}, leaving effectively a free harmonic oscillator with a frequency of the order of inflaton mass at the minimum.} as the one under consideration.

Having a potential that can support oscillon solutions is of course a necessary, although not sufficient, condition for their formation. In order to properly assess the formation of oscillons, we will resort in what follows to non-perturbative $3+1$ classical lattice simulations. We will make use of the recently developed \texttt{$\mathcal{C}osmo\mathcal{L}attice$} package~\cite{Figueroa:2020rrl,Figueroa:2021yhd}. This public code allows to solve the preheating dynamics in a self-consistent way, accounting not only for particle production but also for its backreaction effects on the homogeneous Friedmann equations \eqref{eq:Friedmanneq} and therefore on the scale factor evolution. To this end, it makes use of a cubic box of comoving size $L$, $N^3$ points uniformly distributed and periodic boundary conditions. The values of $L$ and $N$ are chosen in such a way all the relevant momenta involved in our simulations are always well within the infrared (IR) and ultraviolet (UV) resolution in momentum space, 
\begin{equation}
k_{\rm IR}=\frac{2\pi}{L}\,,\hspace{15mm} k_{\rm UV}= \frac{\sqrt{3}}{2} N k_{\rm IR}  \,, 
\end{equation} 
associated respectively with the size of the simulation box and the lattice spacing. 

Following the standard procedure, the initial conditions for the Higgs field are determined by evolving the homogeneous version of Eq.~\eqref{eq:KG-inflaton} from the deep slow-roll regime (${\cal N}_*=55$) till the end of inflation, which we identify with the onset of our simulations ($t=0$). Fluctuations over this homogeneous background are then added as Gaussian random fields, as done usually for systems with short classicalization times as the one under consideration. The subsequent temporal evolution of the system is performed using a 2nd-order Velocity Verlet algorithm, which turns out to be sufficient to keep the violation of energy conservation below $\mathcal{O}(10^{-3})$ during the whole simulation time. More precisely, the scale factor is evolved using the second Friedmann equation 
\begin{equation}\label{Friedmannlattice2}
 \frac{\ddot a}{a} = \frac{1}{3}\langle -2 E_K + E_V \rangle\,,   
\end{equation}
while the first one,
\begin{equation}\label{Friedmannlattice1}
H^2=\frac{1}{3} \langle E_{\rm K}+E_{\rm G}+E_{\rm V}\rangle\,, 
\end{equation}
is rather used as a check of energy conservation. In these expressions, we have conveniently split the total energy density into its kinetic (K), gradient (G) and potential (V) counterparts,
\begin{equation}
E_K\equiv \frac{1}{2}\dot \chi^2  \,, \hspace{10mm} E_G\equiv \frac{1}{2a^2} (\nabla \chi)^2\,,\hspace{10mm} E_V\equiv V \,,
\end{equation}
with $\langle \ldots \rangle$ denoting the spatial averaging of the corresponding quantity over the simulation volume. Note also that the lattice potential accounts for both the large and small field regimes in Eq.~\eqref{eq:potential}, suitably interpolated in our lattice implementation. Whenever needed, we will make also use of temporal averages of specific quantities over a given number of oscillations, denoting them as $\langle \ldots \rangle_T$.

%%%%%%%%%%%%%%%%%%%%%%%%%%%%%%%%%%%%%%%%%%%%%%%%%%%%%%%%%%%%%%%%
\subsection{Oscillon formation}
%%%%%%%%%%%%%%%%%%%%%%%%%%%%%%%%%%%%%%%%%%%%%%%%%%%%%%%%%%%%%%%%

Changing the value of the non-minimal coupling $\xi$ is expected to substantially affect the background dynamics and the associated growth of perturbations, leading to  different fragmentation processes. Indeed, as shown in Fig.~\ref{fig:xi-dependence}, the ratio  $\chi_{\rm end}/\chi_i$, between the field value $\chi_{\rm end}$ at the end of inflation and the location of the inflection point $\chi_i$, increases with $\xi$, allowing the background field to explore the tachyonic region of the potential for a higher number of oscillations. At the same time, the Hubble rate immediately after the end of inflation scales inversely proportional to $\xi$ ($H\sim V^{1/2}_{\rm inf}\sim \sqrt{\lambda}/\xi$), so that higher values of this quantity translate also into a less efficient expansion rate.  
 
To characterize the preheating stage in EC HI and the potential formation of oscillons\footnote{It is important to remark that, contrary to the vast literature on the subject, our localized configurations arise from a potential involving a quartic part, instead of a mere quadratic one, around the minimum. As this does not affect the formation of localized objects, we will refer to them as oscillons.} configurations in this setting, we will consider two benchmark values of $\xi$ leading to different fragmentation modalities. These values are summarized in Table \ref{tab:lattice-parameters}, where we present also the grid parameters, the infrared and ultraviolet resolution of the corresponding reciprocal lattice in units of the oscillation frequency \eqref{eq:rescaling} and the final number of $e$-folds of preheating considered in our simulations.  In both cases, the associated tensor-to-scalar ratio lies substantially below the expected sensitivity of CMB Stage-IV experiments~\cite{Hazumi:2019lys}, making the stochastic GWs background generated during inflation completely undetectable in the near future. 

%%%%%%%%%%%%%%%%%%%%%%%%%%%%%%%%%%%%%%%%%%%%%%%%%%%%%%%%%%%%%%%%%%%%%%%%%%%%%%
\begin{table}
\renewcommand{\tabcolsep}{1.2pc} % enlarge column spacing
\renewcommand{\arraystretch}{1.5} % enlarge line spacing
\hspace{1.5cm}
\begin{tabular}{|c|c|c|c|c|c|} \hline
$\xi$ & $L\left[M^{-1}\right]$ & N & $k_{\rm IR}\left[M\right]$ &  $k_{\rm UV}\left[ M\right]$  & $\Delta \mathcal{N}$ 
\\
\hline
\hline 
$5.0\cdot 10^4 $& 30 & 512 & 0.2 & 89  & 2 
\\
\hline
$2.5\cdot 10^5$ & 60 & 512 & 0.1 & 44  & 1
\\
\hline
\end{tabular}
\caption{Lattice parameters and resolution for the two benchmark scenarios considered in our simulations. For each of these cases, we present also the final number of $e$-folds of preheating considered in our simulations, $\Delta \mathcal{N}$. } \label{tab:lattice-parameters}
\end{table}
%%%%%%%%%%%%%%%%%%%%%%%%%%%%%%%%%%%%%%%%%%%%%%%%%%%%%%%%%%%%%%%%%%%%%%%%%%%%%%

%%%%%%%%%%%%%%%%%%%%%%%%%%%%%%%%%%%%%%%%%%%%%%%%%%%%%%%%%%%%%%%%%%%%%%%%%%%%%%%%
  \begin{figure}
    \begin{center}
            \begin{minipage}[ht]{0.45\linewidth}
            \center{ {\fontsize{14}{16}\selectfont $\xi=5 \cdot 10^4 $}}
        \end{minipage}            
        \vspace{10mm}
        \begin{minipage}[ht]{0.45\linewidth}
            \center{ {\fontsize{14}{16}\selectfont $\xi=2.5 \cdot 10^5 $}}
        \end{minipage}   
        \vspace{10mm}
        \begin{minipage}[ht]{0.45\linewidth}
            \center{\includegraphics[width=\textwidth]{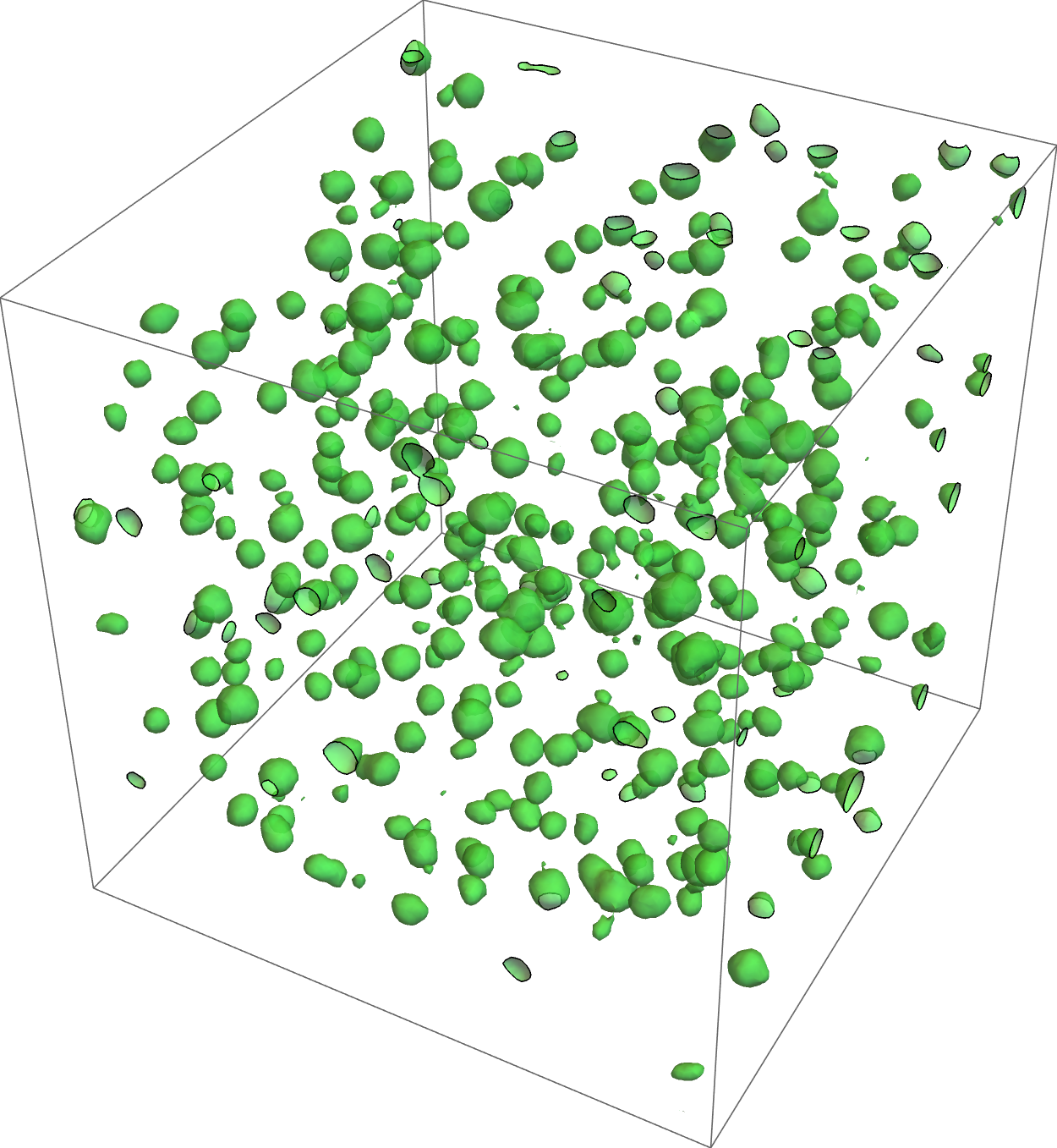} }
        \end{minipage}
                \begin{minipage}[ht]{0.45\linewidth}
            \center{\includegraphics[width=\textwidth]{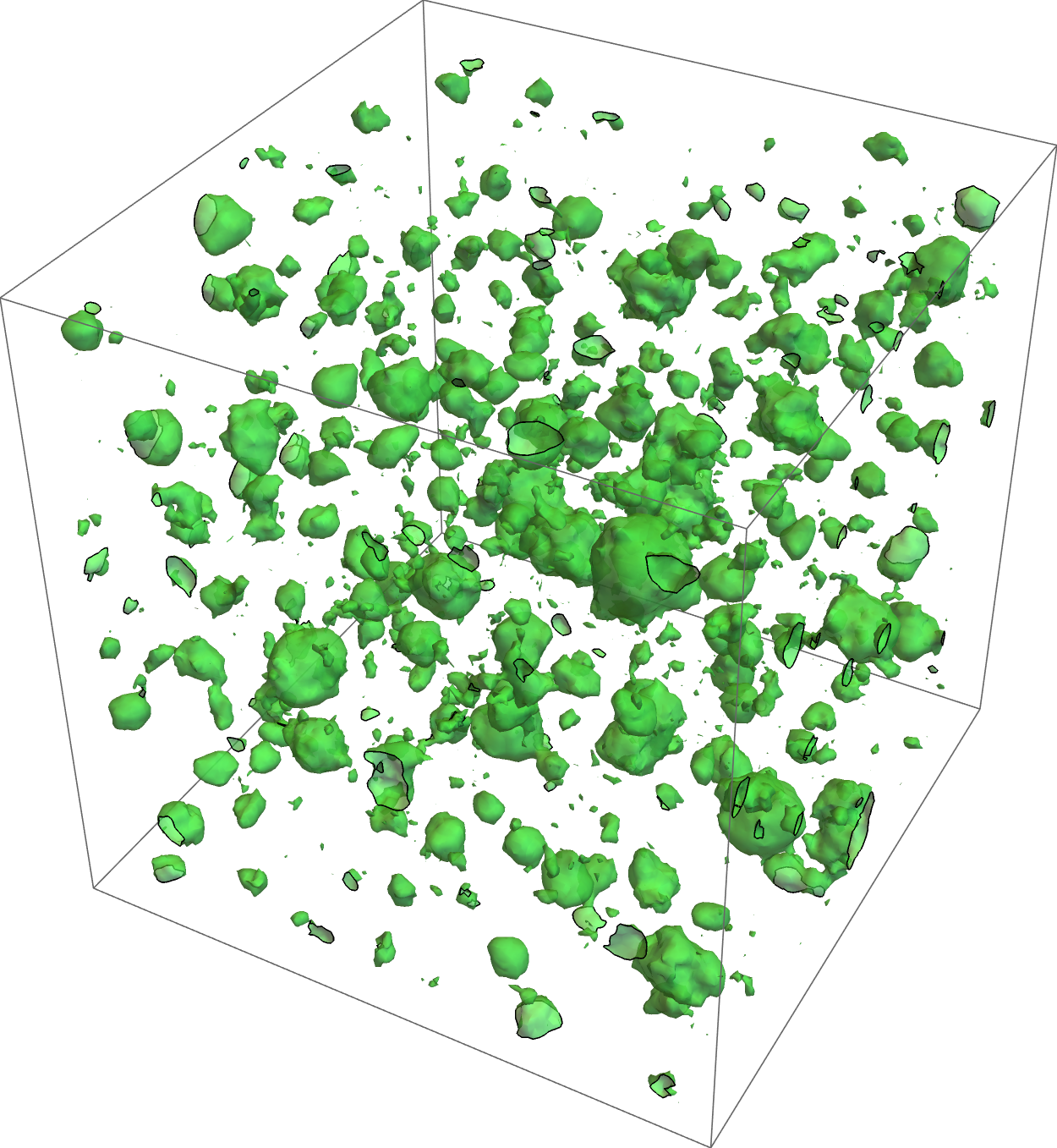} }
        \end{minipage}
    \caption{Oscillons configurations for our two benchmark scenarios at time $Mt = 500$, corresponding respectively to $\Delta {\cal N}=2~({\rm left})$ and $\Delta {\cal N}=0.9~({\rm right})$. 
    The contours in the plots represent regions with overdensities $5$ times larger than the average. Once formed, these objects remain located at almost fixed spatial positions for a large number of oscillations. Due to the expanding box, their fixed physical size shrinks, however, with time.}\label{Fig:snapshots}
    \end{center}
\end{figure}
%%%%%%%%%%%%%%%%%%%%%%%%%%%%%%%%%%%%%%%%%%%%%%%%%%%%%%%%%%%%%%%%%%%%%%%%%%%%%%%%

The typical outputs of our simulations are presented in Fig.~\ref{Fig:snapshots} and in this \href{https://youtu.be/DDwtx44Cfh0}{URL}, where we display several snapshots of the 3-dimensional lattice volume as well as computer-generated movies explicitly demonstrating the formation of oscillons in EC HI. The emergence of these pseudo-solitonic objects manifests itself also through several observables which we now proceed to describe:

 %%%%%%%%%%%%%%%%%%%%%%%%%%%%%%%%%%%%%%%%%%%%%%%%%%%%%%%%%%%%%%%%%%%%%%%%%%%%%%%%
  \begin{figure}
    \begin{center}
        \begin{minipage}[ht]{0.49\linewidth}
            \center{\includegraphics[width=\textwidth]{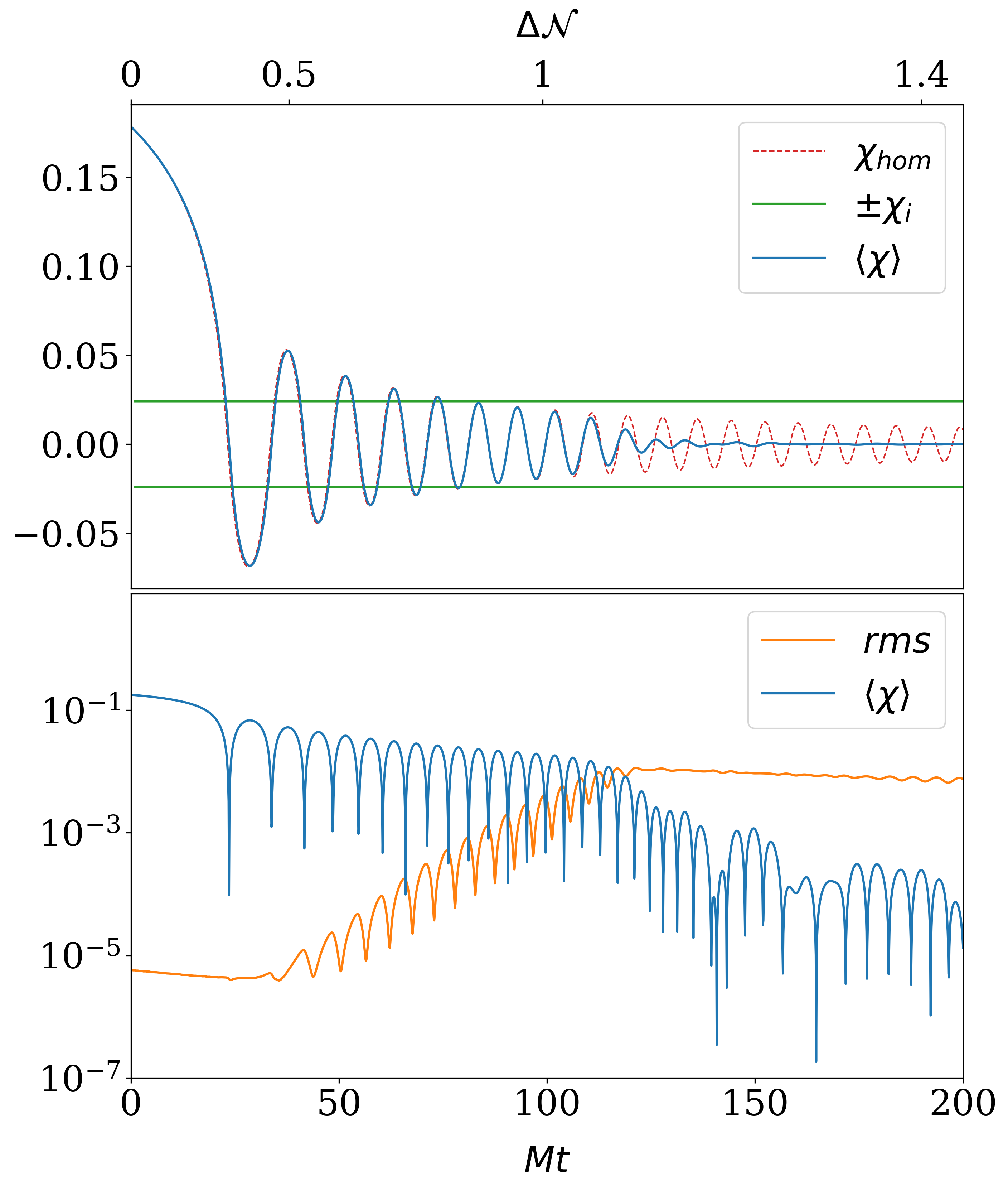} }
        \end{minipage}
                \begin{minipage}[ht]{0.49\linewidth}
            \center{\includegraphics[width=\textwidth]{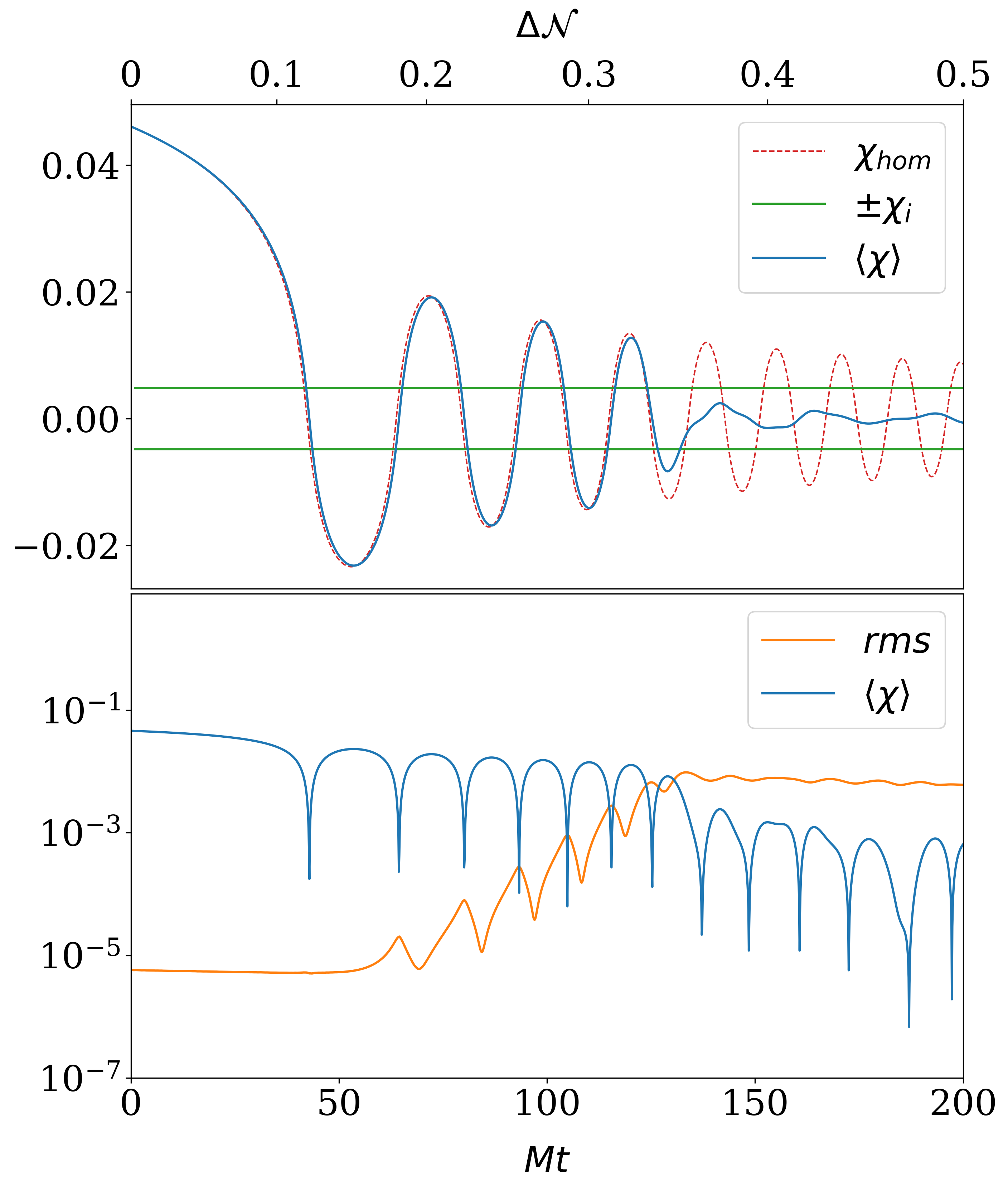} }
        \end{minipage}
    \caption{
    Lattice results for our two benchmark scenarios $\xi=5\cdot 10^4$ (left), $2.5\cdot 10^5$ (right). First row: Time evolution of the lattice-averaged field value $\langle \chi \rangle$ (blue) and the homogeneous solution (red dashed), as compared to the location of the inflection points $\pm \chi_i$ (green). Second row: Time evolution of 
    the root mean squared \eqref{eq:rms} characterizing the growth of perturbation. When this quantity becomes comparable with the amplitude of the background oscillations, fragmentation takes place, and $\langle \chi \rangle$ starts deviating from the homogeneous solution.  }\label{Fig:BG}
    \end{center}
\end{figure}
%%%%%%%%%%%%%%%%%%%%%%%%%%%%%%%%%%%%%%%%%%%%%%%%%%%%%%%%%%%%%%%%%%%%%%%%%%%%%%%%

\begin{itemize}
\item {\bf Field average and time-dependent dispersion:} 
As shown in the upper panels of Fig.~\ref{Fig:BG}, the coherent nature of the inflaton field at sufficiently early times is well-captured by the field average $\langle\chi\rangle$, which undergoes damped oscillations driven essentially by the homogeneous version of Eq.~\eqref{eq:KG-inflaton}. However, the cumulative tachyonic production of Higgs excitations translates eventually into a sharp deviation from this coherent behaviour, with the average field value drastically dropping down to zero at that time. 

As anticipated at the beginning of this section, a change in the non-minimal coupling $\xi$ impacts both the number of times the field undergoes tachyonic oscillations and the overall backreaction dynamics.
 In particular, for $\xi=5\cdot 10^4$ and thanks to the $Z_2$ symmetry of the potential, the homogeneous condensate is able to re-enter the instability region for about consecutive 6-7 semi-oscillations before the cosmic expansion dampens out its amplitude. Afterwards, the field continues oscillating in the quadratic part of the potential, with the condensate remaining intact for about 8-9 additional semi-oscillations to finally fragment in less than one oscillation. On the other hand, for $\xi=2.5 \cdot 10^5$, the Higgs field spends most of the time beyond the inflection point, leading to fragmentation in just 6 semi-oscillations. This situation is reminiscent of what happens in the Palatini formulation of HI, where the field incursions all the way back the inflationary plateau lead to fragmentation in a single oscillation~\cite{Rubio:2019ypq,Dux:2022kuk}. Note, however, that in that case, the potential is essentially quartic all the way up to the plateau, not being able therefore to support the formation of oscillons.

A similar picture follows also from considering the root mean squared (rms) characterizing the typical growth of perturbations,
\begin{equation} \label{eq:rms}
    {\rm rms}(\chi)=\sqrt{\langle \chi^2\rangle -\langle \chi \rangle^2}\,.
\end{equation}
As shown in the lower panels of Fig.~\ref{Fig:BG}, backreaction takes place precisely when this quantity reaches the amplitude of the background component and the average energy density stored in perturbations becomes comparable with that of the inflaton condensate.

%%%%%%%%%%%%%%%%%%%%%%%%%%%%%%%%%%%%%%%%%%%%%%%%%%%%%%%%%%%%%%%%%%%%%%%%%%%%%%
\begin{figure}
\includegraphics[width=0.49\textwidth]{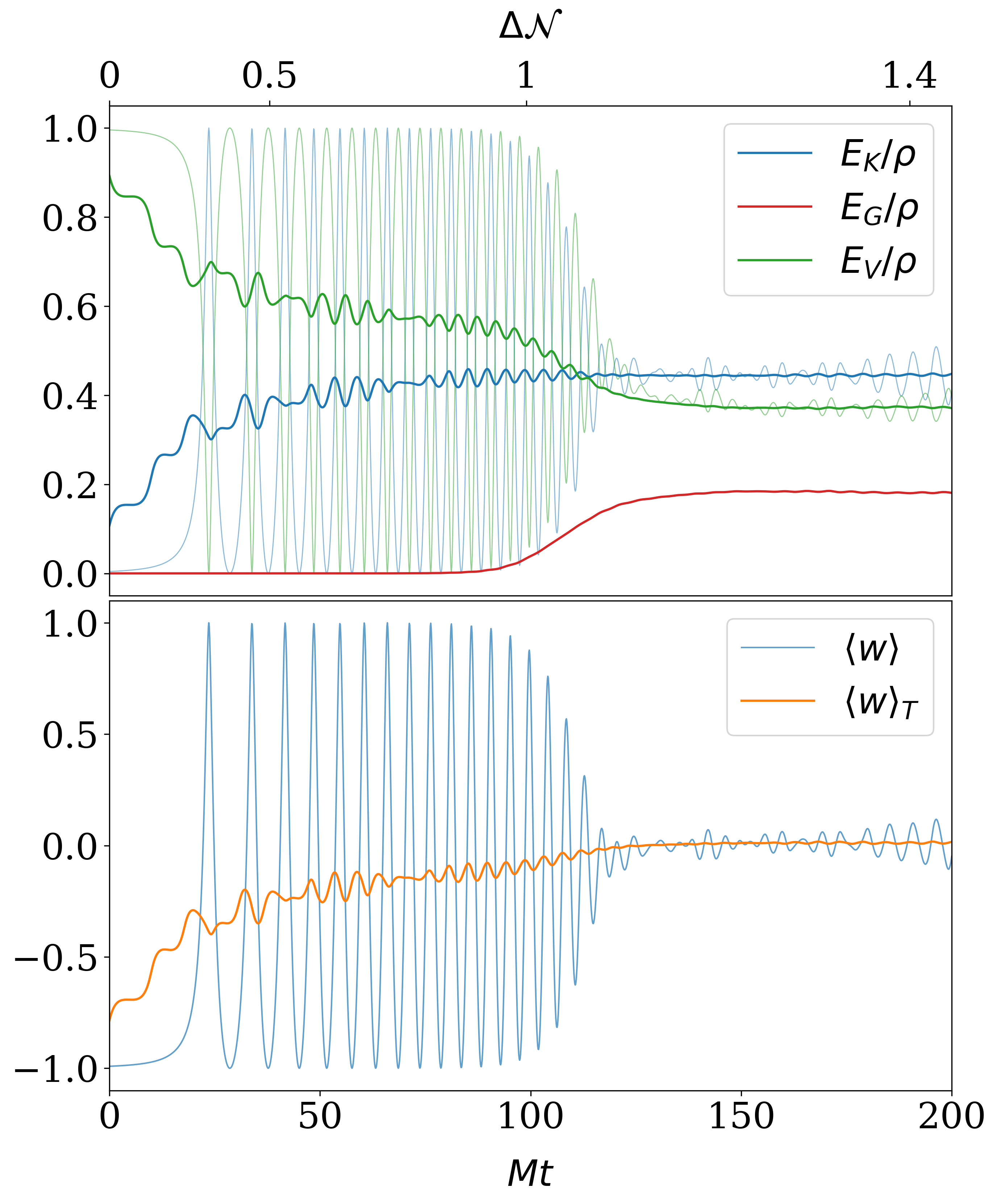} 
\includegraphics[width=0.49\textwidth]{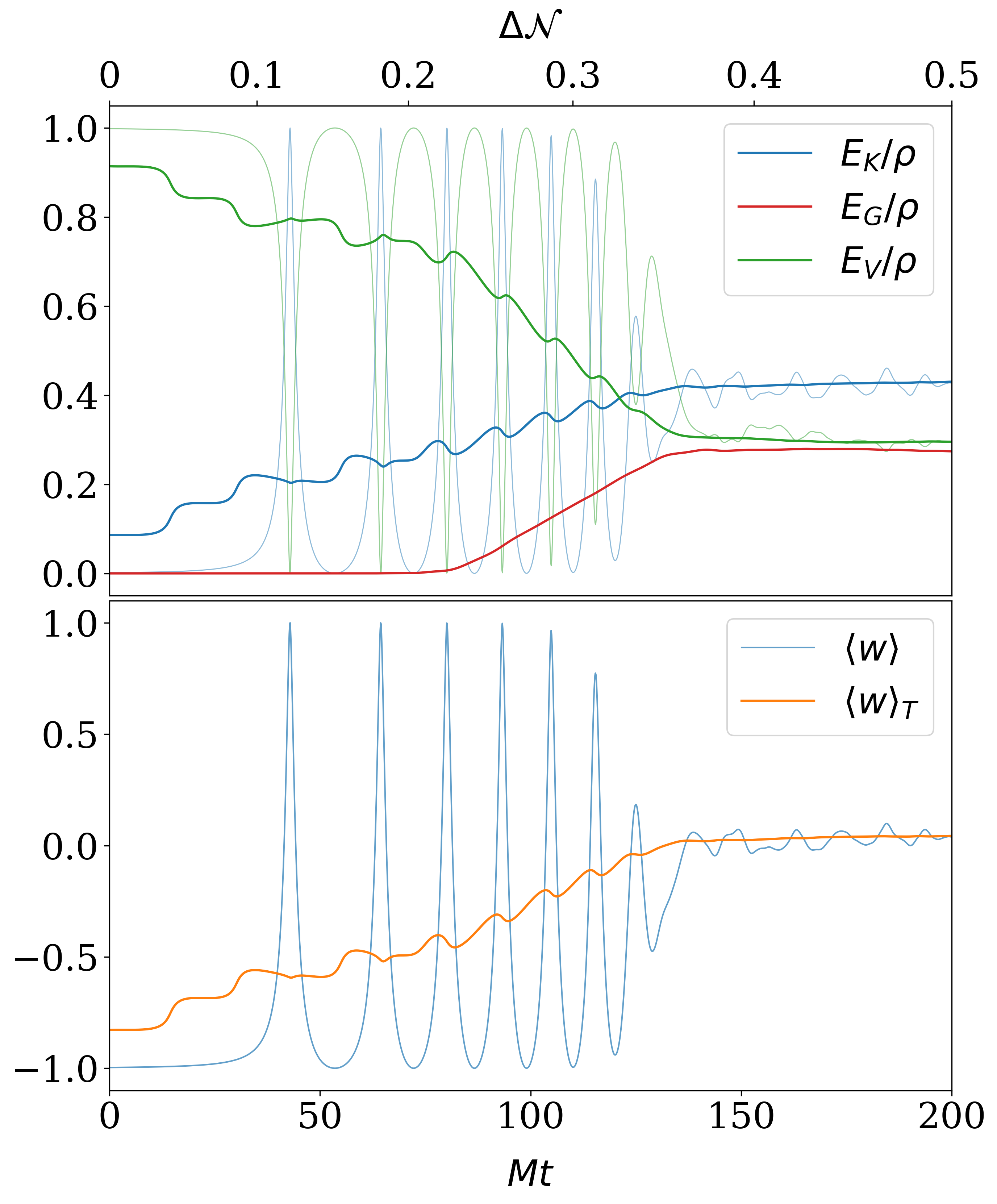} 
    \caption{Lattice results for our two benchmark scenarios $\xi=5\cdot 10^4$ (left), $2.5\cdot 10^5$ (right). First row: Evolution of the volume-averaged (semi-transparent) kinetic (blue), gradient (red) and potential (green) contributions to the total energy, together with their time averages (opaque). Second row: Time evolution of the equation of state parameter (blue) and its time average (orange).  }\label{Fig:energybudget}
\end{figure}
%%%%%%%%%%%%%%%%%%%%%%%%%%%%%%%%%%%%%%%%%%%%%%%%%%%%%%%%%%%%%%%%%%%%%%%%%%%%%%
\item {\bf Energy budget and average equation of state:} 
As previously mentioned, one of the key ingredients allowing the inflaton condensate to re-enter the tachyonic region of the potential is a relatively large ratio between the field value at the end of inflation and the location of the inflection point. However, an important contribution to the dynamics comes also from the expansion rate regulating the damping of oscillations. In the presence of particle creation, this is effectively determined by the relative contribution of kinetic, gradient and potential counterparts to the total energy density, or, equivalently, by the associated equation-of-state parameter,
\begin{equation}
\label{eq:EOS}
w\equiv\frac{\langle p\rangle}{\langle \rho\rangle}=\frac{\langle E_K- E_G/3 -E_V\rangle}{\langle E_K+E_G+E_V\rangle}\,.
\end{equation} 
The evolution of the different energy components for our two benchmark scenarios is displayed in the upper panels of Fig.~\ref{Fig:energybudget}. During the initial tachyonic oscillations, the energy budget is dominated by the potential energy contribution, reflecting the fact that the inflaton field spends more time close to the turning points than to the minimum of the potential. This is also apparent in the evolution of the equation-of-state parameter,  displayed in the lower panels of the same figure. In particular, while this quantity oscillates wildly between $w=-1$ (potential energy domination) and $w=1$ (kinetic energy domination), its temporal average lies always within the interval $-1<\langle w\rangle_T<0$, accounting both for the early-time field excursions in the plateau domain and the light tendency to equipartition between kinetic and potential energy contributions once the oscillation amplitude enters the quadratic part of the potential. This situation holds until the onset of backreaction, where the growth of perturbations leads to the appearance of a non-negligible gradient component that becomes commensurable or even equal to the potential counterpart. At this fragmentation stage, the equation of state approaches rapidly zero, in accordance with Eq.~\eqref{eq:EOS} and the energy budget at that time. The created oscillons behave then collectively as pressureless dust. 

At this point, it is important to remark that, in the parameter range of our simulations, backreaction happens always before the average field amplitude drops below the threshold value $\chi_c$. Therefore, when fragmentation happens and oscillons form, the Universe enters a sustained period of matter domination regardless of the shape of the potential. This contrasts with the naive expectation following from an incomplete homogeneous treatment, where a radiation-dominated stage is expected to develop at field values $\bar{\chi}<\chi_c$, where the EC HI potential \eqref{eq:potential} becomes approximately quartic \cite{Allahverdi:2020bys}.

\item {\bf Power spectra:}  
%%%%%%%%%%%%%%%%%%%%%%%%%%%%%%%%%%%%%%%%%%%%%%%%%%%%%%%%%%%%%%%%%%%%%%%%%%%%%%
\begin{figure}
\begin{center}

\hspace{15mm}\includegraphics[width=0.8\textwidth]{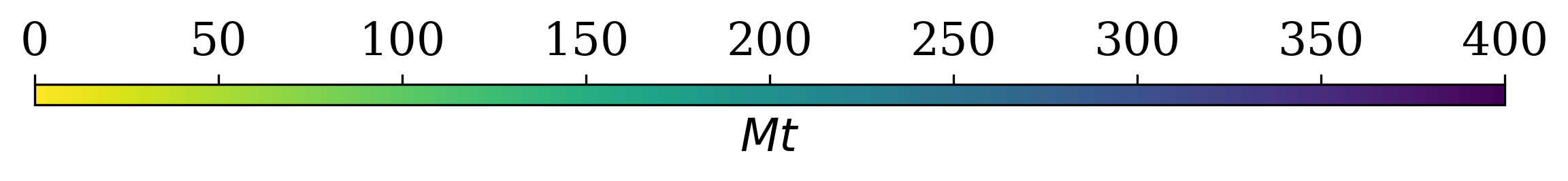}
\includegraphics[width=0.49\textwidth]{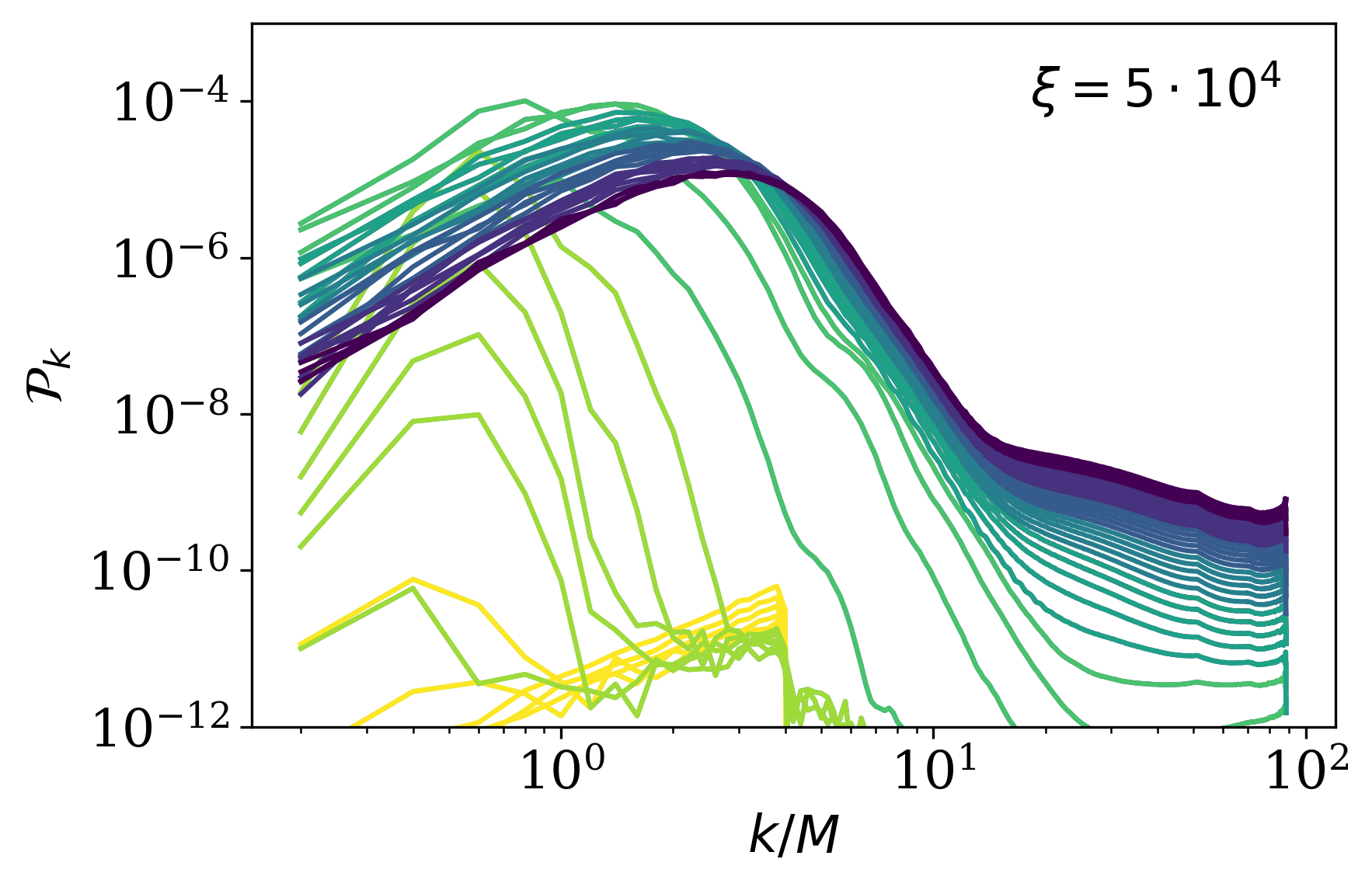} 
\includegraphics[width=0.49\textwidth]{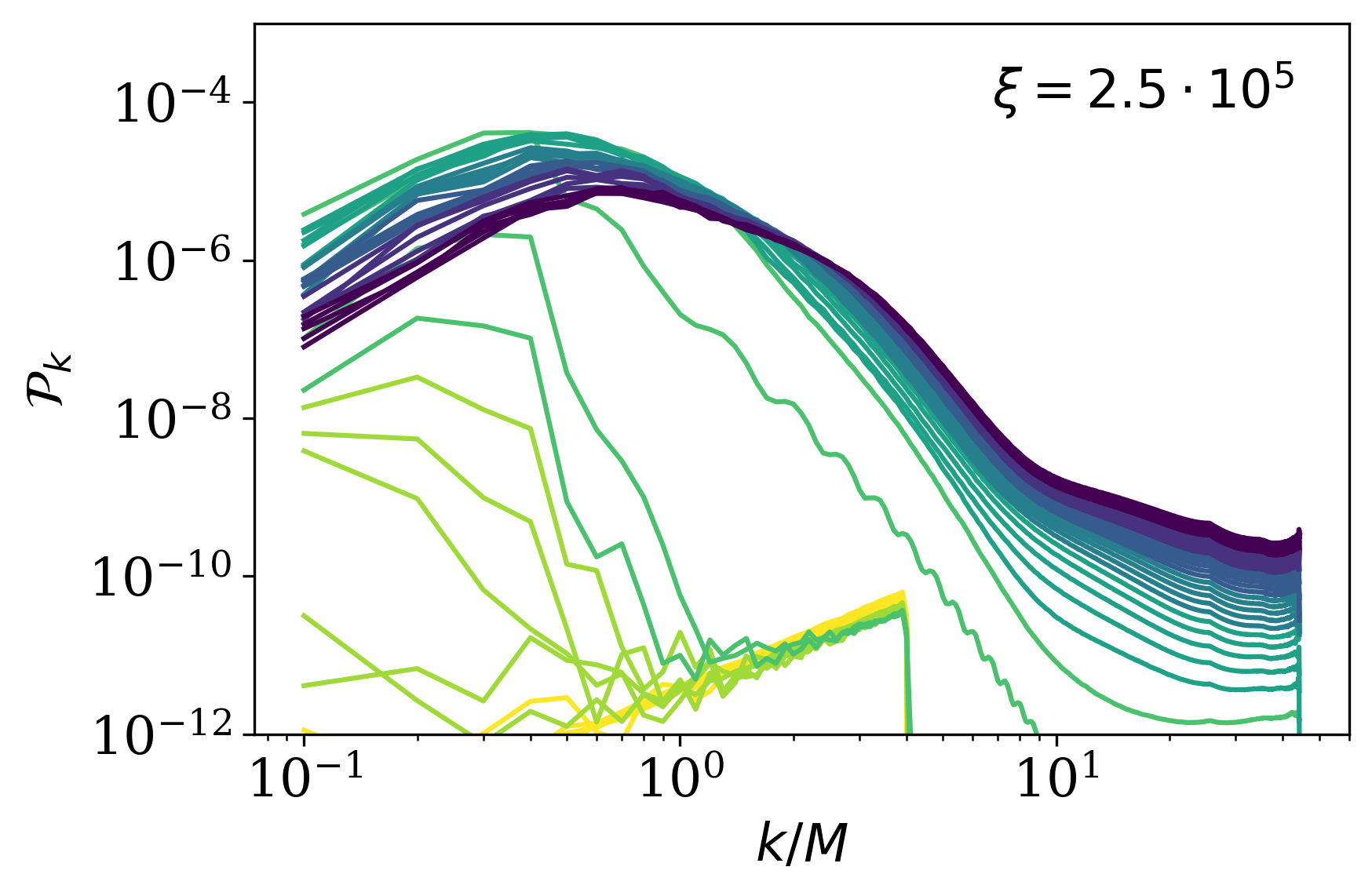} 
\end{center}
    \caption{Time evolution of the energy power spectra for our two benchmark scenarios, with each spectrum taken at intervals of $\Delta (Mt)=10$. The initial growth of perturbations at low momenta is clearly distinguishable, as well as the broad peak appearing due to oscillon formation. Once oscillons have formed, we observe a residual growth of momenta in the UV. This is due to the presence of the quartic region at the bottom of the EC HI potential and does not happen therefore in other purely quadratic models.}\label{Fig:spectra}
\end{figure}
%%%%%%%%%%%%%%%%%%%%%%%%%%%%%%%%%%%%%%%%%%%%%%%%%%%%%%%%%%%%%%%%%%%%%%%%%%%%%%
The unstable tachyonic band in Eq.~\eqref{eq:max-momentum} can be quantitatively observed in the Higgs power spectrum $\mathcal{P}_k$ or two-point correlation function in Fourier space, customarily defined through 
\begin{equation}
    \label{eq:PS-def}
    \langle \chi^2\rangle= \int \dfrac{k^3}{2 \pi^2} d \ln k \,\mathcal{P}_k ~.
\end{equation}
As shown in Fig.~\ref{Fig:spectra}, the associated infrared fluctuations experience a rapid growth at the very early stages of preheating, where the average field value exceeds the location of the inflection point $\chi_i$. As expected, this growth is stronger for higher values of $\xi$, in agreement with the higher efficiency and longer duration of the tachyonic regime in this case.

Due to self-resonance, the enhancement of perturbations continues even for $\langle\chi \rangle< \chi_i$, albeit at a significantly slower rate. Eventually, the growth of fluctuations is halted by backreaction effects and a broad peak appears at a \textit{physical} momentum $\rm k_p$, corresponding to a typical oscillon size $R\sim (2\pi)/{\rm k_p}$.~\footnote{Since the momenta on the lattice are co-moving and the oscillons have a fixed physical size, this scale shift towards the UV due to the expansion of the Universe.} It is interesting to notice that increasing $\xi$ affects also slightly the typical excited band and the final size of the oscillons. As we will see in the forthcoming section, this has a very mild impact on the peak frequency of the produced GWs signal.

\item {\bf Energy density histograms:} The smoking gun that signals the actual existence of quasi-spherical structures of a fixed physical size, rather than other kinds of anisotropies, is the energy density distribution \cite{Lozanov:2019ylm}. As shown in Fig.~\ref{Fig::histograms}, the histograms containing this quantity develop a flattening region at high densities, representing the points on the lattice that lie within the overdense oscillon configurations. This plateau-like feature encodes the local conservation of energy in oscillons as compared to the decreasing energy density of the simulation box $\langle \rho \rangle \propto a^{-3}$, so that the plateau in the distribution becomes more and more pronounced at later times. 

Note that  while the oscillons remain highly subdominant in volume, they dominate the total energy budget. In particular, the fraction of regions where the energy density exceeds the average energy density $\langle \rho \rangle$ by a threshold factor 5, 
\begin{equation}
  f= \frac{\int_{\rho>5\langle 5\rangle} \rho \,dV}{\int\, \rho \, dV}\,,  
\end{equation}
accounts for $\sim 70\%$ of the total energy.
\end{itemize}

%%%%%%%%%%%%%%%%%%%%%%%%%%%%%%%%%%%%%%%%%%%%%%%%%%%%%%%%%%%%%%%%%%%%%%%%%%%%%%%%
\begin{figure}
\includegraphics[width=0.32\textwidth]{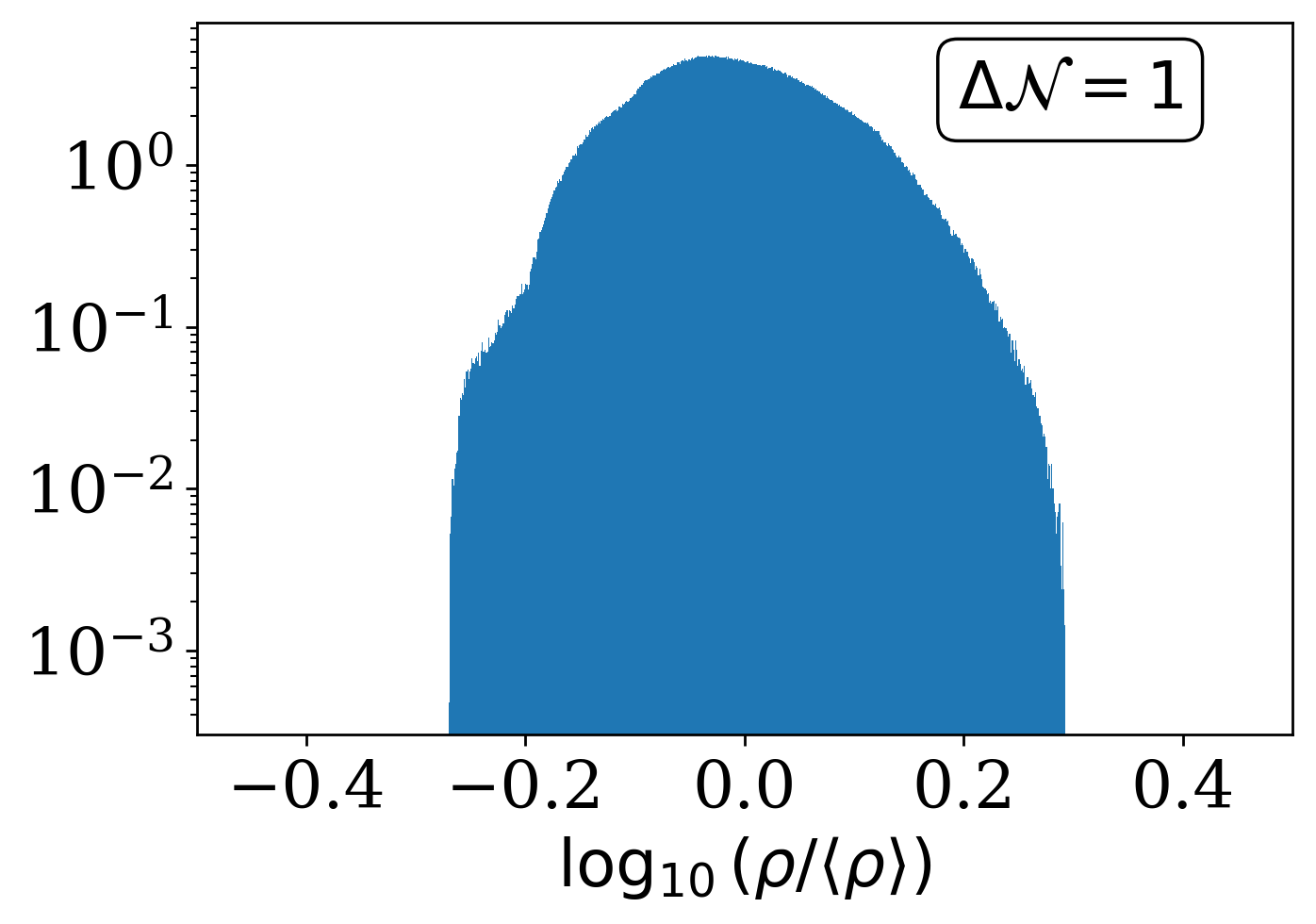} 
\includegraphics[width=0.32\textwidth]{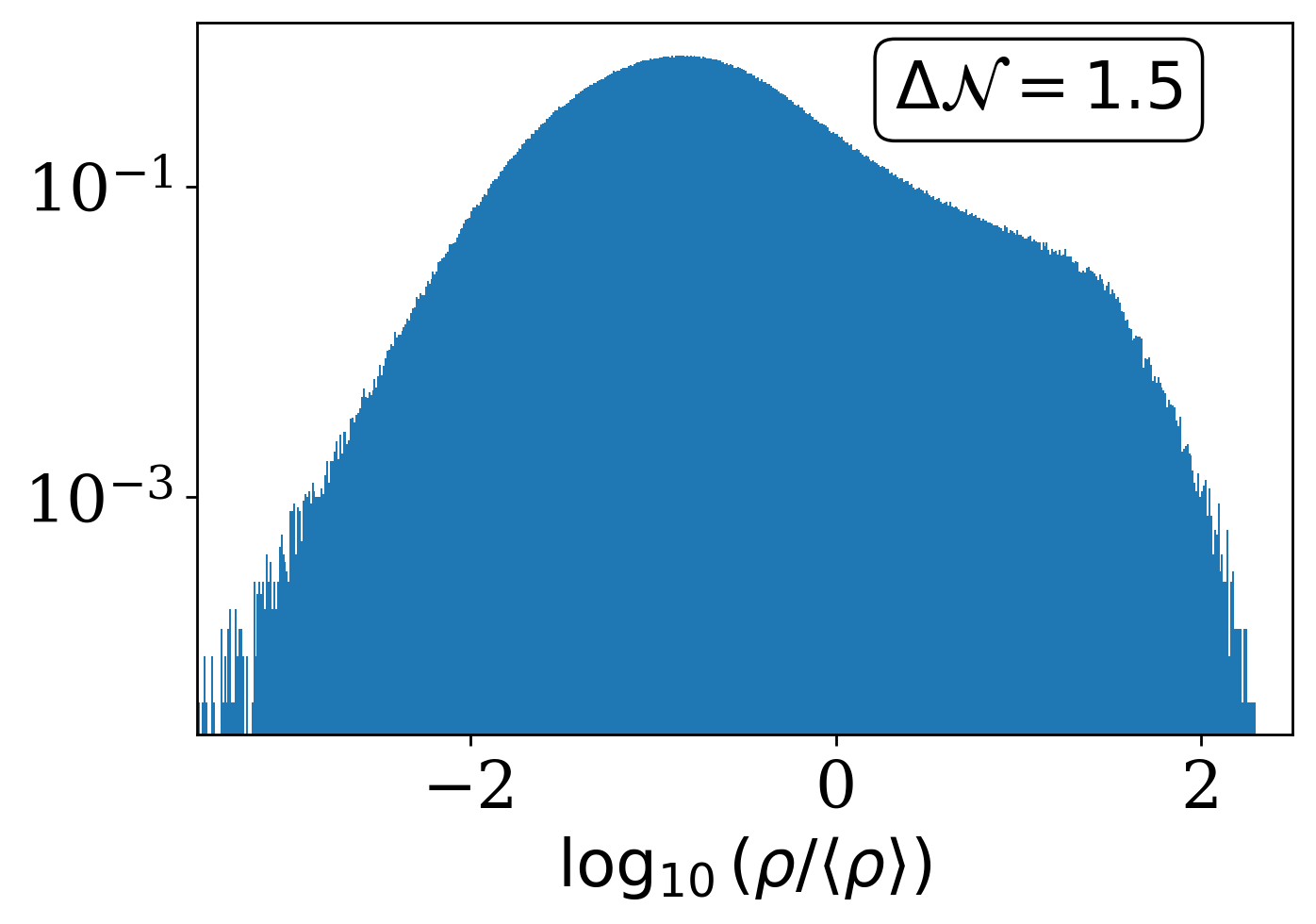} 
\includegraphics[width=0.32\textwidth]{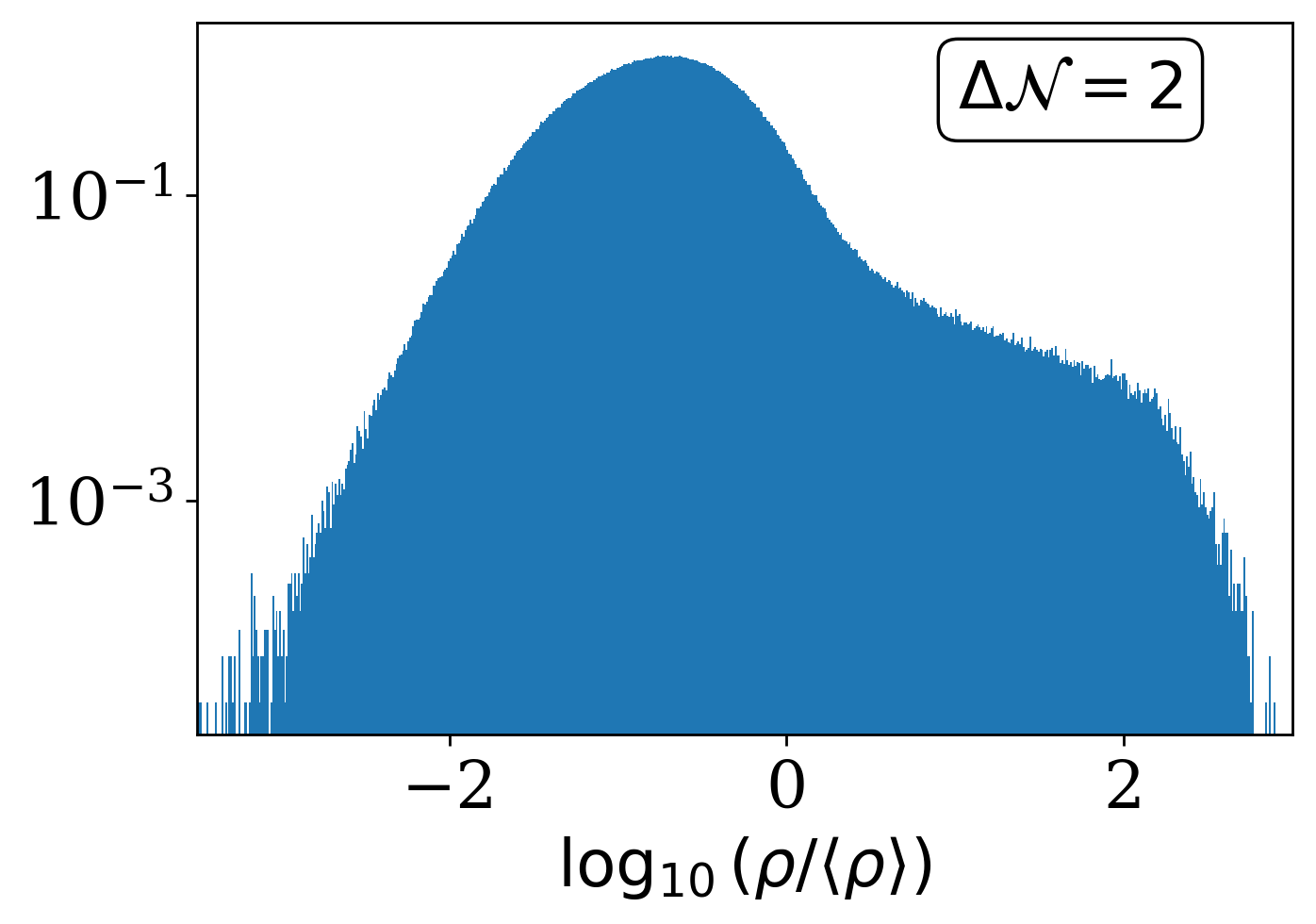} 
\caption{Energy density histograms for the benchmark case $\xi=5 \cdot 10^4$, with the heights of the bins normalized to set the area of the histograms to $1$. One can clearly distinguish the transition from the configuration before fragmentation (left), to the one slightly after oscillon formation (centre), where the distribution starts flattening, signalling the appearance of highly overdense regions with density contrasts $\rho/\langle \rho \rangle >5$. As expected, the flatter part becomes more pronounced at later times (left), as the energy in the oscillons remains constant and the one in the background dilutes as matter. }\label{Fig::histograms}
\end{figure}
%%%%%%%%%%%%%%%%%%%%%%%%%%%%%%%%%%%%%%%%%%%%%%%%%%%%%%%%%%%%%%%%%%%%%%%%%%%%%%%%

%%%%%%%%%%%%%%%%%%%%%%%%%%%%%%%%%%%%%%%%%%%%%%%%%%%%%%%%%%%%%%%%%%%%%%%%%%%%%%%%
\subsection{Gravitational waves signal}\label{sec:3.1.2}
%%%%%%%%%%%%%%%%%%%%%%%%%%%%%%%%%%%%%%%%%%%%%%%%%%%%%%%%%%%%%%%%%%%%%%%%%%%%%%%%

In the previous section, we have seen how oscillons form in EC HI for a wide range of the parameter space compatible with observations, with different values of the non-minimal coupling $\xi$ leading to different fragmentation dynamics. In this section, we explore whether fragmentation and oscillon formation leads to distinctive GWs signatures.

\begin{figure}
\begin{center}
\hspace{15mm}\includegraphics[width=0.8\textwidth]{Pictures/colorbar.png}
\includegraphics[width=0.49\textwidth]{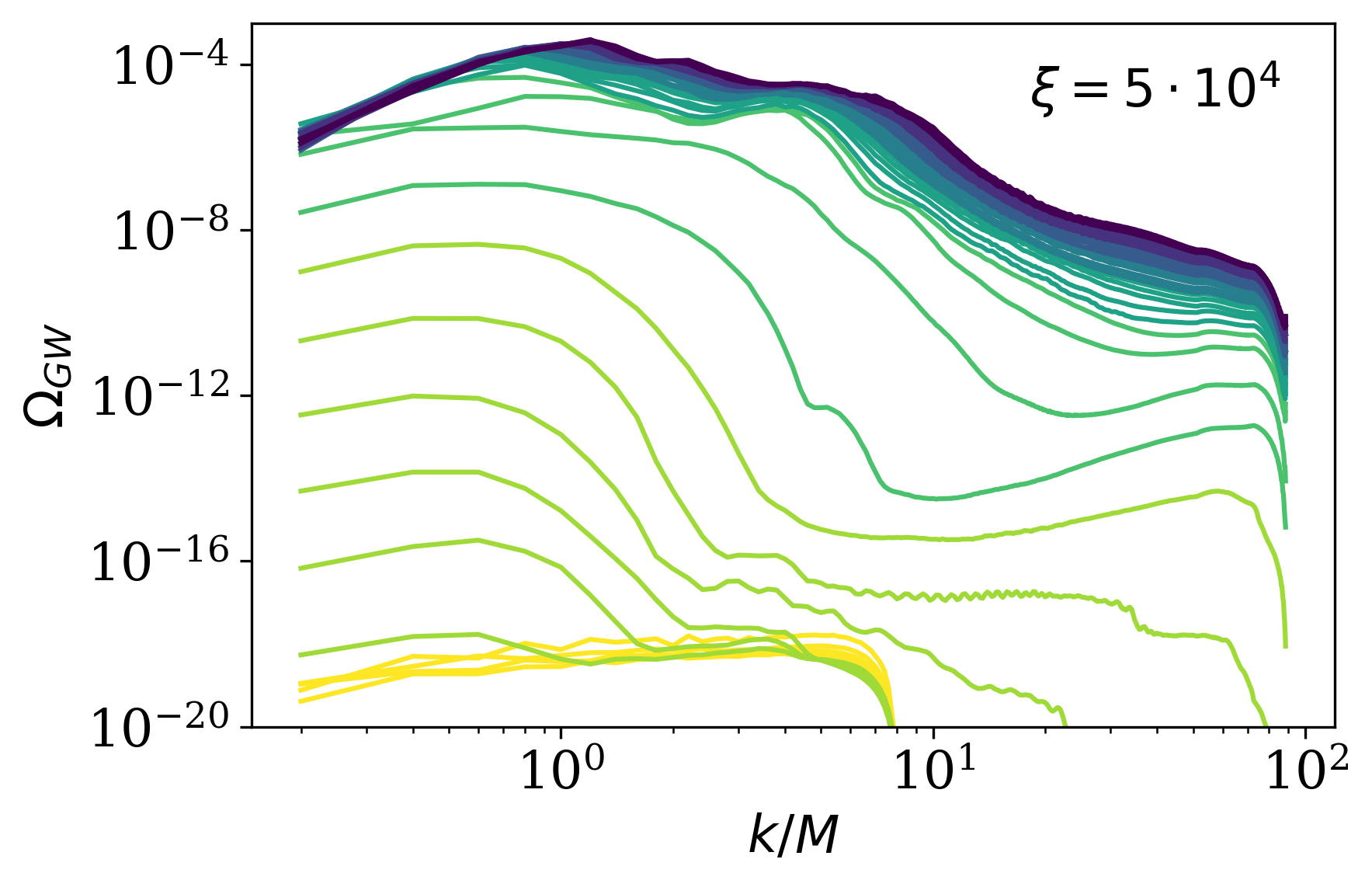} 
\includegraphics[width=0.49\textwidth]{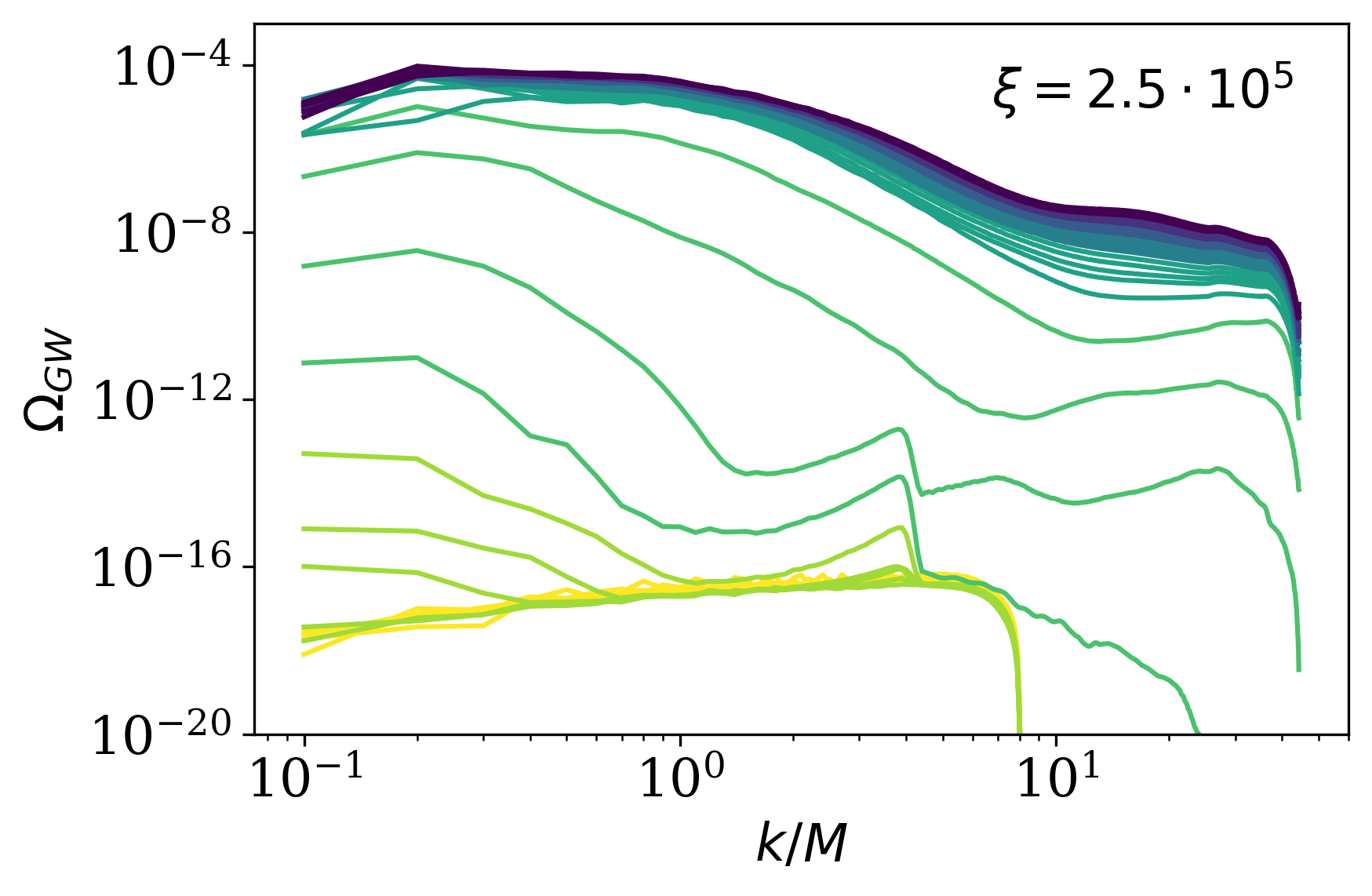}
\end{center}
    \caption{Gravitational waves spectra for our two benchmark scenarios. The first scales to grow are those related to the inhomogeneities arising during the initial growth of perturbations and the fragmentation process. In a second  moment, we observe a secondary peak at higher momenta, due to the oscillon formation and relaxation to a spherical shape. Afterwards, the growth stops, as there are no more strong dynamical inhomogeneities to produce GWs. }\label{Fig:GW-Spectra}
\end{figure}

 The anisotropies in the inflaton configuration are expected to source linear metric perturbation $h_{ij}$ through the wave equation
\begin{equation}
    \label{eq:GW-KG}
    \ddot{h}_{ij}+3 H \dot{h}_{ij}-a^{-2}\nabla^2 h_{ij}=2 a^{-2}\Pi^{\rm TT}_{ij}~,
\end{equation}
with 
\begin{equation}
    \label{eq:PiTT}
    \Pi^{\rm TT}_{ij}=(\partial_i \chi \partial_j \chi)^{\rm TT}
\end{equation}
the transverse-traceless~(TT) part of the associated anisotropic energy-momentum tensor. Employing the GW module implemented in \texttt{$\mathcal{C}osmo\mathcal{L}attice$}~\cite{tech}, we can easily extract the energy density power spectrum of the GWs signal, namely \cite{Figueroa:2020rrl} 
\begin{equation}
    \label{eq:GWPS}
    \Omega_{\rm GW}=\frac{1}{3 H^2} \frac{{\rm d}\rho_{\rm GW}}{{\rm d}\log \,k}~, \hspace{10mm}{\rm with}\hspace{10mm} \rho_{\rm GW}=\frac{1}{4}\langle \dot{h}_{ij} \dot{h}_{ij}\rangle~.
\end{equation}
The results of the simulations for the two benchmark scenarios considered in the previous section are shown in Fig.~\ref{Fig:GW-Spectra}. We observe that the GWs spectrum grows initially with time, developing a main peak at low momenta. When fragmentation starts, the spectrum quickly rises, and, shortly after fragmentation ends, there appears a subdominant secondary peak at a higher frequency. Once the oscillons have formed the spectrum stops growing, reflecting their individual character and roughly spherical symmetry. 
%%%%%%%%%%%%%%%%%%%%%%%%%%%%%%%%%%%%%%%%%%%%%%%%%%%%%%%%%%%%%%%%%%%%%%%%%%%%%%%%

%%%%%%%%%%%%%%%%%%%%%%%%%%%%%%%%%%%%%%%%%%%%%%%%%%%%%%%%%%%%%%%%%%%%%%%%%%%%%%%%

The proper connection of the GW spectrum at the time of emission to that at present times requires the knowledge of the full cosmological history upon fragmentation, and in particular, of the average lifetime of the created oscillon configurations. Unfortunately, observing the fate of these very stable objects within a realistic scenario like the one under consideration requires not only extensive computational power but also a detailed lattice implementation of the Higgs couplings to other Standard Model degrees of freedom, a task that significantly exceeds the proof-of-principle scope of this paper (cf., however, Section \ref{sec:4}). In order to estimate the current amplitude $\Omega_{0,{\rm GW}}$ and frequency $f_0$ of the spectrum, we follow therefore a poor-man's parametric approach based on two working approximations. First, we will presume the Universe to evolve adiabatically until the present day, ensuring entropy conservation. Second, we will consider that, once oscillons have formed, the associated equation-of-state parameter evolves gradually to that of radiation domination, the moment at which we assume the SM matter content to be completely thermalized.~\footnote{Note that, although most likely true in a SM setting like the one under consideration \cite{Bezrukov:2014ipa,Kurkela:2011ti}, the onset of radiation domination does not necessarily coincide with that of thermalization. In particular, the former requires only the effective energy dominance of relativistic particles, which might have, however, a non-thermal spectrum, see e.g.~Refs.~\cite{Micha:2002ey,Micha:2004bv}.} With these assumptions, we get ~\cite{Dufaux:2007pt}
\begin{eqnarray} \label{eq:peak-frequency}
&& f_0=\frac{k_p}{2\pi a_0}=\dfrac{ k_{p}}{a_{\rm e} \rho_{\rm e}^{1/4}}e^{{-\frac{1}{4}(1-3\bar w)\Delta {\cal N}_{\rm RD} }}\, 4 \cdot 10^{10} Hz~, \\
&& \Omega_{0,{\rm GW}}~h^2 \simeq 1.6 \times 10^{-5} e^{-(1-3\bar w)\Delta {\cal N}_{RD}} \, \Omega_{\rm e,GW}\,,
\end{eqnarray}
with $a_0$ the scale factor today, the quantities $a_{\rm e}$, $\rho_{\rm e}$ and $\Omega_{\rm e,GW}$ denoting respectively the scale factor, the total energy density and the gravitational waves' energy fraction at the time of emission and $\bar w$ the mean equation-of-state parameter in the $e$-folds interval ${\cal N}_{\rm RD}$ between GW emission and the onset of radiation domination. 
As explicit in Eq.~\eqref{eq:peak-frequency}, there are two main factors contributing to the redshift of the maximum available signal. On the one hand, the likely existence of a prolonged intermediate stage with an equation of state $0<\bar w<1/3$. On the other hand,  $f_0$ will be smaller for higher values of the energy density at the time of formation. Nonetheless, a higher energy density during preheating implies also a smaller horizon and hence higher momenta in the spectrum, so that opposite contributions from $k_p$ and $\rho_e$ tend to cancel out. 
As usually happens in preheating scenarios, the resulting amplitude of the signal is commensurable with the sensitivity of current GWs experiments, but the peak frequency turns out to be ${\cal O}(\rm GHz)$, lying therefore far away from the available observational window\footnote{See~\cite{Aggarwal:2020olq} for experimental proposals in the $\rm MHz-GhZ$ window.}.
As a consequence, the main distinctive feature of the EC formulation, compared to the Palatini and metric one, is the possibility of a prolonged stage of matter domination, with consequences on the inflationary observables.

%%%%%%%%%%%%%%%%%%%%%%%%%%%%%%%%%%%%%%%%%%%%%%%%%%%%%%%%%%%%%%%%%%%%%%%%%%%%%%%
\section{Towards preheating in the Standard Model}\label{sec:4}
%%%%%%%%%%%%%%%%%%%%%%%%%%%%%%%%%%%%%%%%%%%%%%%%%%%%%%%%%%%%%%%%%%%%%%%%%%%%%%%

Now that we have established the presence of oscillons in the scalar sector of EC-HI, let us turn our attention to the other SM components, namely the massive electroweak bosons and fermions. As shown in Refs.~\cite{Bezrukov:2008ut,Garcia-Bellido:2008ycs,Bezrukov:2014ipa,Repond:2016sol,Rubio:2019ypq}, the generation of these two species is intrinsically intertwined, with the Bose enhancement of the former strongly depending on their perturbative decay into fermions. 

The so-called \textit{Combined preheating formalism} developed in Ref.~\cite{Garcia-Bellido:2008ycs} (cf. also Ref.~\cite{Bezrukov:2014ipa}) allows for a semi-analytic treatment of the gauge boson production in Higgs inflation, under several simplifying assumptions. First, it replaces the usual Higgs-gauge boson interactions following from the SM covariant derivatives with global couplings between the Higgs field and three scalar degrees of freedom playing the role of gauge bosons. This approximation is well-supported by several numerical studies, explicitly demonstrating the qualitative equivalence of global and Abelian interactions during the initial stages of preheating, where the non-linearities associated with gauge boson self-interactions can be still safely neglected \cite{Figueroa:2015rqa,Enqvist:2014tta,Enqvist:2015sua}. Second, the formalism makes use of a WKB approximation in order to determine the production of gauge boson excitations at every semi-oscillation of the inflaton field. This well-established procedure takes into account the spontaneous and stimulated emission of particles at the bottom of the potential, as well as potential interference effects among semi-oscillations \cite{Kofman:1997yn}. Third, \textit{Combined preheating} accounts for the potential decay of the produced gauge bosons into the SM quarks and leptons by considering effective decay widths proportional to the gauge boson masses, as done, for instance, in instant preheating scenarios \cite{Felder:1998vq}. This approximation explicitly neglects a plethora of annihilations processes, which are, however, subdominant at early times or low gauge boson number densities \cite{Bezrukov:2014ipa}. All together the \textit{Combined preheating} assumptions give rise to a master,
phase-averaged, equation relating the number densities $n_k(j)$ of each gauge boson field polarization at two consecutive zero crossings \cite{Bezrukov:2014ipa}
\begin{equation}\label{eq:depletion}
\left[\frac{1}{2}+n_k((j+1)^+)\right]\hspace{-1mm}= (1+2 C_k(j))\left[\frac{1}{2}+n_k(j^+)\,e^{-\Theta(j)}\right]\,,
\end{equation}
 with $C_k(j)$ a $k$-dependent IR window function depending on the small-amplitude behaviour of the corresponding gauge boson mass, 
\begin{equation}
\Theta(j)=\int_{t_{j}}^{t_j+1} \Gamma(t) d t
\end{equation}
the mean decay rate into fermions among two consecutive crossings $t_{j-1}$ and $t_j$ and $\Gamma(t)$ the instantaneous rate evaluated on the background equations of motion. The explicit form of these building functions can be obtained by transforming the usual (\textit{scalarized}) Higgs-gauge boson interactions, $\sim g_i^2/8 \, h^2 A^2$, into the Einstein frame, with $g_i=g, g/\cos \,\theta_w$ for the $A=W^\pm$ and $Z$ bosons, $g$ the $U(1)_Y$ coupling and $\theta_w$ the weak mixing angle. Upon performing an additional field redefinition $A\to \Omega A$  for canonically normalizing the resulting kinetic sector, we obtain an interaction term\footnote{For simplicity we will always refer to quantities related to the $W^\pm$ bosons, Although the same formulas, with the appropriate couplings, apply to the $Z$ boson.}
\begin{equation}
\frac{{\cal L}_{int}}{\sqrt{-g}}=  \frac{1}{2}m^2A^2\,,
\end{equation}
with 
\begin{equation}
    \label{eq:Einstein-GB}
    m^2=\dfrac{g^2}{4}\dfrac{h^2}{1+\xi h^2}=\begin{cases} 
\dfrac{g^2}{4}\dfrac{\chi^2}{1+\xi \chi^2}~, &  \hspace{10mm}   \chi<\chi_c~,  \\
\dfrac{g^2}{4 \xi}\left[1-\exp\left(- \dfrac{2\xi|\chi|}{\sqrt{c}}\right) \right]~,& \hspace{10mm} \chi \gg\chi_c\,,
   \end{cases}
\end{equation}
an effective mass approaching the constant value $g^2/(4\xi)$ in the large field regime and dropping to zero at each minimum-crossing of the inflaton field, where the adiabaticity condition ${\dot{m}/m^2<1}$ is maximally violated and gauge boson production takes place. In this latter regime, the equation of motion for the $A$-field perturbations takes the form  
\begin{equation}
    \label{eq:daughter-perturbation}
    \ddot{\delta A_{\textbf{k}}}+3 H\dot{\delta A_{\textbf{k}}} +\left( \textbf{k}^2/a^2 +m^2(\chi)\right)\delta A_{\textbf{k}}=0~.
\end{equation}
with 
\begin{equation}
    \label{eq:linearized}
m^2(\chi) \simeq \begin{cases} 
\dfrac{g^2}{4} \chi^2~, &  \hspace{5mm} \textrm{for} \hspace{5mm} \chi<\chi_c~,  \\
\dfrac{g^2}{2\sqrt{c}}|\chi|~,& \hspace{5mm} \textrm{for} \hspace{5mm} \chi \gg\chi_c~,
   \end{cases}
\end{equation} 
depending only on the non-minimal coupling $\xi$ through the constant $c$, which, for a given value of $\lambda$, is almost completely determined by the CMB normalization, cf.~Eq.~\eqref{eq:inflationary-parameters}. 
%%%%%%%%%%%%%%%%%%%%%%%%%%%%%%%%%%%%%%%%%%%%%%%%%%%%%%%%%%%%%%%%%%%%%%%%%%%%%%%%%

\begin{figure}
\begin{center}
\includegraphics[width=0.4\textwidth]{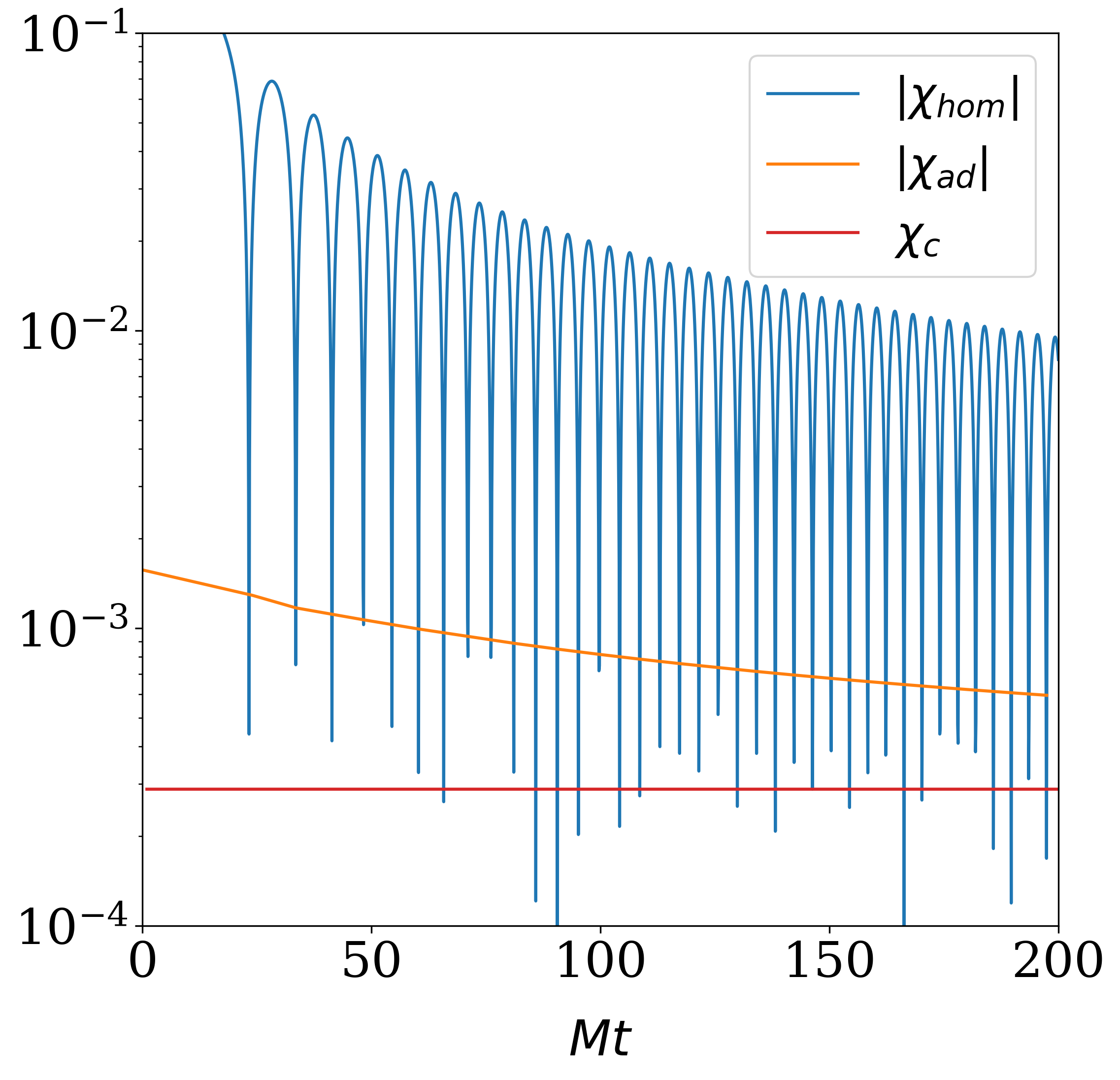} 
\includegraphics[width=0.4\textwidth]{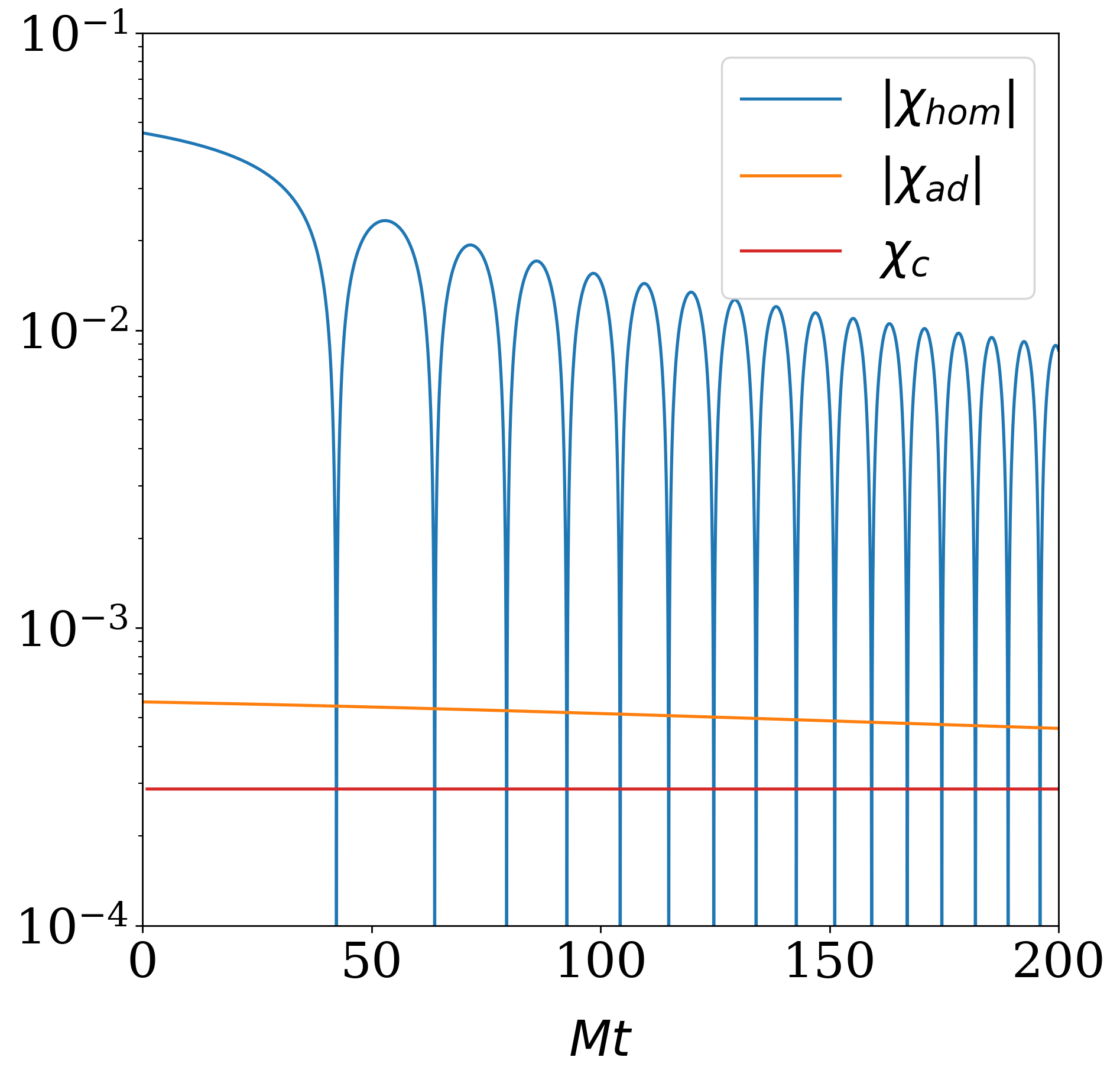} 
    \caption{Field values at which adiabaticity condition ${\dot{m}/m^2<1}$ is first violated in our two benchmark scenarios (orange), as compared to the homogeneous solution for the background (blue) and the crossover value $\chi_c$ below which the Einstein-frame potential becomes quartic (red). Note that, although adiabaticity is always broken in the quadratic part of the potential, the value $\chi_{ad}$ approaches $\chi_c$ at higher $\xi$.}\label{Fig:Adiabaticity-breaking}
\end{center}
\end{figure}

\begin{figure}
\begin{center}
\includegraphics[width=0.8\textwidth]{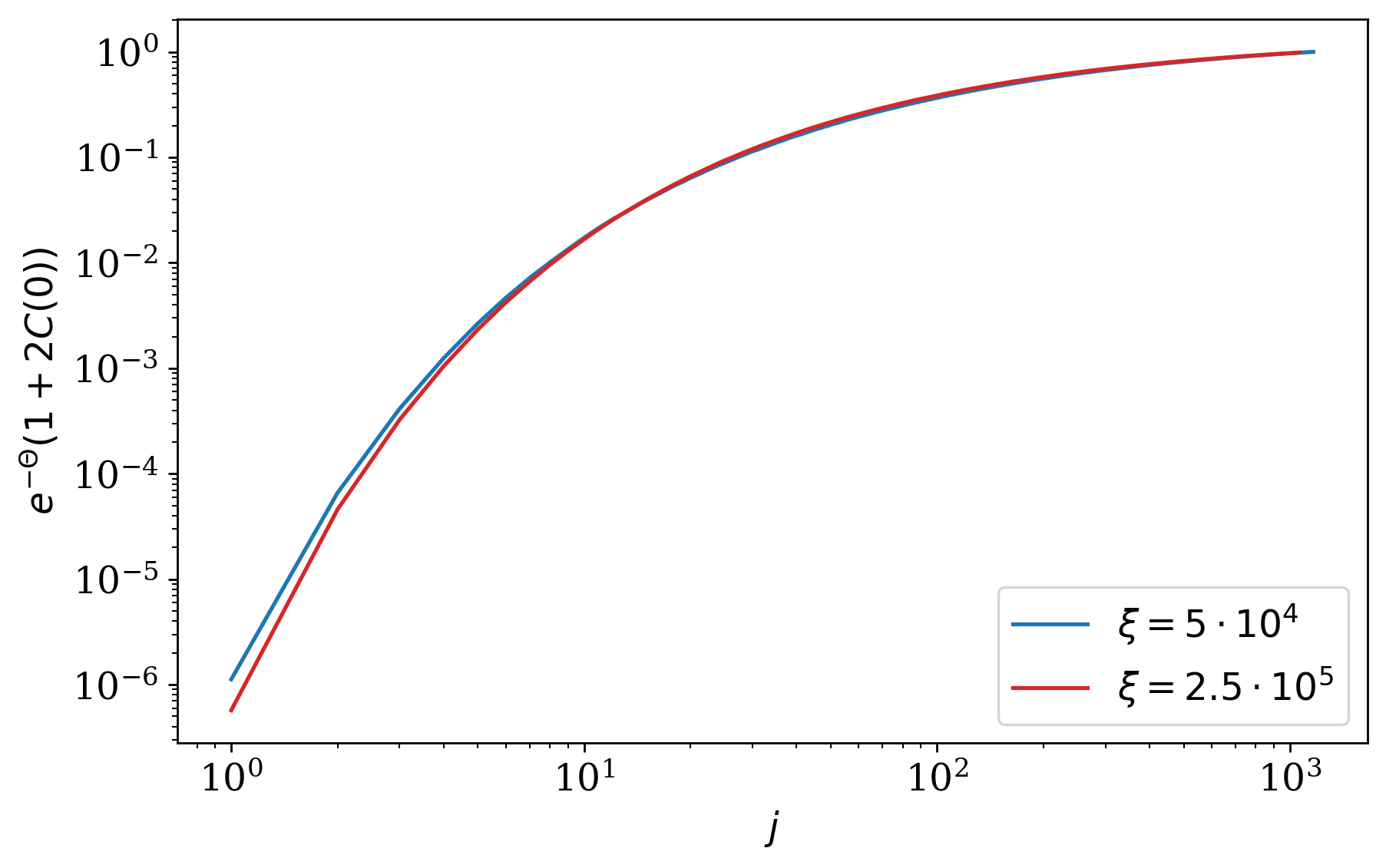} 
    \caption{Interplay between the Bose enhancement factor for $k=0$ and the suppression due to gauge boson decay into fermions for a fiducial gauge coupling $g^2=0.3$, as a function of the number of zero-crossings $j$. The factor $\Theta$ is computed on the numerical solution for the background. Although the two effects cancel out after $\mathcal{O}(10^3)$ zero-crossings, the actual modes can start growing slightly before, once the proper recursive formula in Eq.~\eqref{eq:depletion} is taken into account. Nonetheless, we expect the gauge boson production to be heavily subdominant during the first tens of oscillations where fragmentation and oscillon formation takes place. } \label{Fig:decay}
\end{center}
\end{figure}

%%%%%%%%%%%%%%%%%%%%%%%%%%%%%%%%%%%%%%%%%%%%%%%%%%%%%%%%%%%%%%%%%%%%%%%%%%%%%%%%%
As shown in Fig.~\ref{Fig:Adiabaticity-breaking}, the violation of adiabaticity condition $\Dot{m}/m^2<1$ following from the numerical solution of the background equations of motion happens generically at field values $\chi_a>\chi_c$, even though these two scales approach each other for increasing values of $\xi$. 
Regarding possible perturbative decays into fermions and gauge bosons, there are two key conditions that must be satisfied at the same time. On one hand, for the decay to be kinematically allowed, the condition $M>2m(2 m_f)$ must be fulfilled. On the other hand, the decay can take place efficiently only if the rate $\Gamma$ is larger than the Hubble rate $H$. While for the $W^\pm$ and $Z$ bosons the first condition is never fulfilled, the Higgs can still decay into all the fermions, except the top quark, with different situations. As the decay rate is proportional to the fermion mass, the Higgs will be more likely to decay into the heavier fermions. Nonetheless, these decay channels remain kinematically blocked for timescales much longer than the one for oscillon formation. On the contrary, the decay into light fermions will be kinematically open at early times, but the smallness of their Yukawa couplings, and therefore decay rates, will make the decay inefficient on timescales much longer than the ones relevant to preheating. 
Taking this into account, together with the fact that we only aim at providing a qualitative picture of this otherwise complicated dynamics, we will neglect in what follows the contribution of the upper term in the right-hand side of Eq.~\eqref{eq:linearized}. In this limit, the scenario under consideration reduces conceptually to that studied in the metric formulation of the theory~\cite{Bezrukov:2008ut,Garcia-Bellido:2008ycs,Bezrukov:2014ipa,Repond:2016sol}. By performing the same type of computations, the IR window function in Eq.~\eqref{eq:depletion} takes the form\footnote{For the sake of simplicity, we neglect the slight differences among $W^\pm$ and $Z$ encoded in Lorentz invariant phase space factors, cf.~Ref.~\cite{Garcia-Bellido:2008ycs} for details on this issue. }  \cite{Garcia-Bellido:2008ycs} 
%%%%%
%%%%%
\begin{equation}\label{eq:c_j}
C(x)=\pi^2\left[ {\rm Ai}(-x^2){\rm Ai}'(-x^2)+{\rm Bi}(-x^2){\rm Bi}'(-x^2)\right]^2\,,
\end{equation} 
 with $\rm Ai$ and $\rm Bi$ the Airy functions of first and second kind, $x=k/a M \left(j/q\right)^{1/3}$, $j$ the number of inflaton zero crossings and the parameter
\begin{equation}
q=\frac{\sqrt{c}g^2 \chi_{\rm end}}{8 \pi \lambda}
=\frac{c g^2}{16 \pi\lambda\,\xi } \log \left(1+\frac{2 \sqrt{2} \xi }{\sqrt{c}}\right)\,,
\end{equation}
decreasing for increasing values of $\xi$. Assuming that the same approximation holds in our case, gauge bosons will be produced with a typical scale $k_{\rm gb}\sim q^{\frac{1}{3}} M$, which for our benchmark scenarios $\xi=5\cdot 10^4$ and $2.5 \cdot 10^5$, corresponds roughly to $k_{\rm gb}/M \sim 20,10$, respectively. 
The depletion of the bosons produced at a crossing is efficient whenever the factor $\Theta$ in Eq.~\eqref{eq:depletion} is large enough to compensate for the growth of $n_k$ due to parametric resonance.
Taking into account that the specific form of the decay rate into SM particles is proportional to the mass of the bosons,
\begin{equation}
     \Gamma=\dfrac{3 g^2 m}{16 \pi}\,,
\end{equation}
we can compare the two effects for different values of $\xi$.
 On the one hand, being the maximum decay rate proportional to the gauge bosons masses, and the latter to $1/\sqrt{\xi}$, we have that for larger values of the non-minimal coupling, the maximum $\Gamma$ will have a lower value. On the other hand, increasing $\xi$ substantially modifies the background dynamics, resulting in a higher period for the first few oscillations. 
To get an estimate of the actual timescales on which the decay rate becomes inefficient, we can consider the maximum value of the Bose enhancement factor happening at $x=0$, and compare it with the suppression factor $e^{-\Theta}$, by evaluating the decay rate on the homogeneous solution for the background. As shown in Fig.~\ref{Fig:decay}, in both our scenarios the decay rate is highly efficient for the first $\mathcal{O}(10^2-10^3)$ zero crossings, significantly exceeding the typical number of semi-oscillations needed for oscillon formation, $\mathcal{O}(10-20)$.
%%%%%%%%%%%%%%%%%%%%%%%%%%%%%%%%%%%%%%%%%%%%%%%%%%%%%%%%%%%%%%%%%%%%%%%%%%%%%%%%
%%%%%%%%%%%%%%%%%%%%%%%%%%%%%%%%%%%%%%%%%%%%%%%%%%%%%%%%%%%%%%%%%%%%%%%%%%%%%%%%
This results in two main effects. Firstly, for a large number of oscillations, almost all the energy stored in the gauge bosons at each zero crossing is transferred to SM fermions, making this the dominant channel for energy transfer to SM particles other than the Higgs. Secondly, the SM gauge bosons cannot efficiently accumulate, so their parametric resonance is substantially delayed, preventing them from backreacting on the inflaton condensate, already at the early stages of its evolution.

Now that we have established that the decay into fermion efficiently suppresses the production of gauge bosons, it is natural to wonder how the interplay between the inflaton and other SM particles can affect the formation of oscillons. Based on our simple linear analysis it is clear that the amount of energy transferred to fermions through the gauge bosons decay needs several oscillations to become important. For instance, in the metric case~\cite{Repond:2016sol}, this can take up to hundreds of oscillations. Taking into account that increasing the non-minimal coupling reduces the number of oscillations before fragmentation, we can infer that such transfer becomes less efficient for larger $\xi$, leaving the process of oscillon formation unaltered. However, it is important to point out that our previous analysis works only at a linear level, and does not take into account possible non-linear effects associated with the quick tachyonic growth of inflaton perturbations, as well as the highly inhomogeneous state between fragmentation and oscillon formation. Although it would be interesting to perform a fully non-linear analysis, including both fermions and gauge bosons in our lattice, such a task presents some technical difficulties. First, although \texttt{$\mathcal{C}osmo\mathcal{L}attice$} can in principle support the implementation of the full Standard Model gauge group, this can only be done in the Jordan-Frame, where the covariant derivatives are linear in the fields. Second, even if the gauge fields are introduced as mere scalar fields \cite{Repond:2016sol,Dux:2022kuk}, the created particles will be produced with typical momenta much higher than those relevant for oscillons, therefore requiring simulations with very high resolution, far beyond the scope of this paper. In particular, the coverage of all momenta involved in the picture, especially at later times, would require at least $N>1024$ lattice sites for realistic gauge couplings $g^2\simeq 0.3$. Such simulations would be extremely time-consuming, especially if all three gauge bosons are taken into account.~\footnote{Using a $N=512$ lattice, we are able to confirm, however, that the impact of gauge bosons in the tachyonic amplification of Higgs excitations remains subdominant until oscillon formation for $g^2\simeq {\cal O}(10^{-2})$, even if the decay into fermions is completely neglected and the depletion mechanism discussed above is absent.} Third, when introducing spinors on the lattice one encounters the infamous \textit{fermion-doubling problem} \cite{Nielsen:1981hk}, leading to the appearance of additional copies of the latter. Although this limitation could be circumvented with an effective treatment along the lines of Ref.~\cite{Repond:2016sol} (see also Ref.~\cite{Fan:2021otj}), the resolution of the IR and UV momenta involved would still remain.

%%%%%%%%%%%%%%%%%%%%%%%%%%%%%%%%%%%%%%%%%%%%%%%%%%%%%%%%%%%%%%%%%%%%%%%%%%%%%%%%
\section{Conclusions and outlook}\label{sec:conclusions}
%%%%%%%%%%%%%%%%%%%%%%%%%%%%%%%%%%%%%%%%%%%%%%%%%%%%%%%%%%%%%%%%%%%%%%%%%%%%%%%%

Among the plethora of inflationary models in the literature, Higgs inflation stands out as one rather minimalistic scenario not requiring the introduction of additional degrees of freedom beyond the Standard Model content while potentially allowing for a complete analysis of the preheating stage following the inflation. In spite of its simplicity, the inclusion of a non-minimal coupling of the Higgs field to gravity breaks the usual degeneracy among metric-affine representations, with different formulations of gravity giving rise to different inflationary predictions and preheating dynamics. 

In this paper, we have studied the preheating phase for Higgs Inflation in Einstein-Cartan gravity using fully-fledged numerical lattice simulations in $3 + 1$ dimensions. We have found that this scenario allows for the formation of dense and spatially localized oscillon configurations, constituting up to 70\% of the total energy density for suitable model parameters. These pseudo-solitonic objects survive for the entire simulation time, leading to a prolonged period of matter domination that modifies the post-inflationary history and therefore the minimum duration of the inflationary phase needed to solve the hot Big Bang problems. Furthermore, we have shown that such structures can source a significant gravitational wave signal, providing an alternative observational channel for this inflationary model, besides the usual stochastic background of primordial tensor perturbations generated during inflation. Unfortunately, as usually happens in preheating scenarios, the associated peak frequency turns out to be significantly larger than the one accessible by current and planned  GWs experiments.

While the results of this paper have been explicitly derived assuming a simple Nieh-Yan interaction \eqref{eq:Nieh-Yan}, they can be easily extended to the more general constant $c$ scenarios within the EC multiverse\footnote{For instance, a choice $\xi_{vv}=\xi_{aa}=\xi$, $c_{av}=0$ in Eq.~\eqref{eq:cgeneric} translates also into a constant $c$ value,
\begin{equation}
c= \xi+6 \xi_{\eta,{\rm eff}}^2 \hspace{10mm} \textrm{with} \hspace{10mm} \xi_{\eta,{\rm eff}}^2 \equiv  \xi ^2+ \frac{2}{3}\left(\frac{ \zeta_{ha}^2}{c_{aa}}+\frac{ \zeta_{hv}^2}{c_{vv}}\right)\,,
\end{equation}
allowing to recover the phenomenology described in this paper for any set of parameters $\zeta_{ha}$, $\zeta_{hv}$, $c_{aa}$, $c_{vv}$ leading numerically to $\xi_{\eta,{\rm eff}}^2\simeq \xi_\eta^2$ and commensurable $\xi$ values.} \cite{Karananas:2021zkl}, including also scale-invariant extensions \cite{Piani:2022gon} or even TDiff generalizations \cite{Casas:2018fum,Blas:2011ac}. 
Indeed, the main condition for oscillon formation is effectively encoded in the field redefinition $d\bar \chi/d\bar h$ entering Eq.~\eqref{Hrate} or, equivalently, in the kinetic function $K(h)$ appearing in the Einstein frame-action \eqref{eq:EF-Action}. Provided this quantity is sufficiently close to one (but still larger), the inflaton field will be allowed to exceed the inflection of the potential for a given number of oscillations, building up fluctuations through tachyonic instabilities and eventually fragmenting out into oscillon configurations. 

Our results are only a first step towards the full characterization of the preheating process in Einstein-Cartan HI. In particular, it would be interesting to address several aspects not properly taken into account in this study:
\begin{enumerate}

\item \textit{Oscillons lifetime}: Understanding the lifetime of oscillons is indeed of uttermost importance for determining the precise value of the inflationary observables to be confronted with observations. On top of that, oscillons could also generate gravitational waves when they decay, with longer lifetimes translating into higher chances of observing GWs within the frequency windows of current and future gravitational wave experiments.
In all of our simulations, it is clear that oscillons have enough time to form before the quartic coupling becomes relevant. However, their long-time classical and quantum stability \cite{Segur:1987mg,Gleiser:2009ys,Hertzberg:2010yz,Salmi:2012ta,Evslin:2023qbv}, as well as their possible formation for higher values of $\xi$, remains to be checked. Unfortunately, performing simulations for these scenarios turns out to be rather problematic, as the quartic potential affects mainly the momenta in the far UV regime, while the typical scale of oscillons is located in the IR. In this regard, it would be interesting to test whether the presence of quartic self-interactions can lead to parametric resonance effects within oscillons, along the lines of Ref.~\cite{Hertzberg:2010yz}. Future research could also explore the impact of the complete SM structure on the inflaton lifetime, accounting for rescattering effects induced by non-Abelian interactions \cite{Enqvist:2014tta,Enqvist:2015sua} or potential spike effects in longitudinal gauge degrees of freedom \cite{DeCross:2015uza,DeCross:2016fdz,DeCross:2016cbs,Ema:2016dny,Sfakianakis:2018lzf}. Note, however, that the Riemannian spikes advocated in the latter references are expected to be subdominant in our context. In particular, the combination of the uncertainty principle and the approximate conservation of energy during the first few oscillations, translates roughly into a typical energy scale $\Delta E_{\rm sp}\sim \Delta t_{\rm sp}^{-1}\sim \dot \chi_{\rm 0}/\chi_c \sim  \sqrt{c\,V_{\rm end}}\sim \beta \sqrt{\lambda}$ for the particles created by the spike, with $ \Delta t_{\rm sp}$ the time needed to cross the spike, $\chi_0$ the field velocity at the minimum of the potential, $V_{\rm end}$ the potential energy at the end of inflation and  
\begin{equation}
\beta\equiv \frac{\sqrt{c}}{\sqrt{c}+2\sqrt{2}\xi}\,.     
\end{equation}
While $\beta\sim {\cal O}(1)$ in the metric formulation \cite{Ema:2016dny}, this quantity is significantly  smaller in the EC HI settings
considered in this paper, $\beta\sim {\cal O}(10^{-2}-10^{-3})$, leading to a transfer of energy through this mechanism of order $\rho_{\rm sp}/V_{\rm end}\sim 1/(8\pi^2) (\Delta E_{\rm sp})^4/V_{\rm end}\sim \lambda\, c \, \beta^2/(16\pi^2)\sim {\cal O}(10^{-2}-10^{-3})$.

\item \textit{Gravitational perturbations and primordial black holes}: Since oscillons represent well-localized objects with high overdensities, it is natural to wonder whether these quasi-spherical structures could potentially collapse into black holes, a possibility not accounted for in our simulations, where metric perturbations are completely disregarded. Assuming spherical symmetry, the strength of the Newtonian potential at the surface of a single oscillon of radius $R=(2\pi/{\rm k}_p)$, mass $\mathcal{M}$, and core energy density $\rho_c$ can be estimated as\footnote{This result comes from the fact that oscillons are expected to have a field amplitude at the core proportional to that of the oscillating homogeneous condensate at the time of backreaction, $\chi_c\sim\chi_{\text{br}}$, which result in $\rho_c\sim\langle \rho\rangle_{\text{br}}$.} \cite{Lozanov:2019ylm}
\begin{equation}
    \label{eq:newtonian potential}
    |\Phi|\sim \dfrac{\mathcal{M}}{8\pi R}\sim \dfrac{\rho_c R^2}{6}\sim H^2_{\text{br}}R^2~,
\end{equation}
with $H_{\text{br}}$ the value of the Hubble parameter at the time of backreaction (fragmentation). For all the scenarios considered in this paper, this \textit{compactness} turns out to be significantly smaller than one, $|\Phi_{\rm EC}|\sim 10^{-3}-10^{-4}$, supporting our approximations. 
We point out, however, that this estimate addresses only the typical oscillon sizes that we expect to find in our simulation, being a priori possible, although unlikely due to the magnitude of $\Phi$, that some of them meet the condition for collapse. Moreover, if stable enough, oscillons could eventually cluster on long time scales \cite{Cotner:2018vug,Amin:2019ums}.  
\item \textit{Fermions in EC gravity}: As mentioned when introducing EC gravity, one of the main perks of this formulation is the possibility to naturally account for the coupling of fermions to gravity, thanks to the presence of a non-vanishing spin connection. However, the use of a theory with torsion leads inevitably to the appearance of higher-order four-fermion and scalar-fermion interactions appearing when torsion is integrated out \cite{Kibble:1961ba}. Interestingly enough, these operators can be used, for instance, to produce singlet fermions that can play the role of dark matter, as done in~Ref.~\cite{Shaposhnikov:2020aen}. Unfortunately, even if the universality of gravitational couplings to fermions is then assumed, one would need to take into account potential non-canonical kinetic terms for fermions, increasing the freedom of the model, and reducing, therefore, its predictive power. As long as the scale introduced by these operators is higher than the inflationary one, we can expect their effect to be subdominant both in the running of $\lambda$, as well as in the gauge bosons' perturbative decay. Nonetheless, they might play an important role if non-perturbative effects are taken into account. 
\end{enumerate}
We leave the in-depth study of these interesting aspects for future work. 

%%%%%%%%%%%%%%%%%%%%%%%%%%%%%%%%%%%%%%%%%%%%%%%%%%%%%%%%%%%%%%%%%%%%%%%%%%%%%%%%
\acknowledgments
%%%%%%%%%%%%%%%%%%%%%%%%%%%%%%%%%%%%%%%%%%%%%%f%%%%%%%%%%%%%%%%%%%%%%%%%%%%%%%%%%

 M.~P. (ORCID ID 0000-0002-2387-5948) acknowledges the Funda\c c\~ao para a Ci\^encia e a Tecnologia (FCT), Portugal, for the financial support to the Center for Astrophysics and Gravitation-CENTRA, Instituto Superior T\'ecnico,  Universidade de Lisboa, through the Project No.~UIDB/00099/2020.  M.~P. thanks  also the support of this agency through the Grant No. SFRH/BD/151003/2021 in the framework of the Doctoral Program IDPASC-Portugal. JR (ORCID ID 0000-0001-7545-1533) is supported by a Ram\'on y Cajal contract of the Spanish Ministry of Science and Innovation with Ref.~RYC2020-028870-I. The authors acknowledge the Department of Theoretical Physics of the Universidad Complutense de Madrid for allowing access to the cluster facilities utilized for the research reported in this paper.
%%%%%%%%%%%%%%%%%%%%%%%%%%%%%%%%%%%%%%%%%%%%%%%%%%%%%%%%%%%%%%%%%%%%%%%%%%%%%%%%

%%%%%%%%%%%%%%%%%%%%%%%%%%%%%%%%%%%%%%%%%%%%%%%%%%%%%%%%%%%%%%%%%%%%%%%%%%%%%%%%
\bibliographystyle{JHEP}
\bibliography{mybib}

\providecommand{\href}[2]{#2}\begingroup\raggedright\begin{thebibliography}{100}

\bibitem{Planck:2018jri}
{\scshape Planck} collaboration, \emph{{Planck 2018 results. X. Constraints on inflation}}, \href{https://doi.org/10.1051/0004-6361/201833887}{\emph{Astron. Astrophys.} {\bfseries 641} (2020) A10} [\href{https://arxiv.org/abs/1807.06211}{{\ttfamily 1807.06211}}].

\bibitem{BICEP:2021xfz}
{\scshape BICEP, Keck} collaboration, \emph{{Improved Constraints on Primordial Gravitational Waves using Planck, WMAP, and BICEP/Keck Observations through the 2018 Observing Season}}, \href{https://doi.org/10.1103/PhysRevLett.127.151301}{\emph{Phys. Rev. Lett.} {\bfseries 127} (2021) 151301} [\href{https://arxiv.org/abs/2110.00483}{{\ttfamily 2110.00483}}].

\bibitem{Guth:1980zm}
A.H.~Guth, \emph{{The Inflationary Universe: A Possible Solution to the Horizon and Flatness Problems}}, \href{https://doi.org/10.1103/PhysRevD.23.347}{\emph{Phys. Rev. D} {\bfseries 23} (1981) 347}.

\bibitem{Linde:1981mu}
A.D.~Linde, \emph{{A New Inflationary Universe Scenario: A Possible Solution of the Horizon, Flatness, Homogeneity, Isotropy and Primordial Monopole Problems}}, \href{https://doi.org/10.1016/0370-2693(82)91219-9}{\emph{Phys. Lett. B} {\bfseries 108} (1982) 389}.

\bibitem{Mukhanov:1981xt}
V.F.~Mukhanov and G.V.~Chibisov, \emph{{Quantum Fluctuations and a Nonsingular Universe}}, {\emph{JETP Lett.} {\bfseries 33} (1981) 532}.

\bibitem{Kallosh:2019hzo}
R.~Kallosh and A.~Linde, \emph{{CMB targets after the latest Planck data release}}, \href{https://doi.org/10.1103/PhysRevD.100.123523}{\emph{Phys. Rev. D} {\bfseries 100} (2019) 123523} [\href{https://arxiv.org/abs/1909.04687}{{\ttfamily 1909.04687}}].

\bibitem{Bassett:2005xm}
B.A.~Bassett, S.~Tsujikawa and D.~Wands, \emph{{Inflation dynamics and reheating}}, \href{https://doi.org/10.1103/RevModPhys.78.537}{\emph{Rev. Mod. Phys.} {\bfseries 78} (2006) 537} [\href{https://arxiv.org/abs/astro-ph/0507632}{{\ttfamily astro-ph/0507632}}].

\bibitem{Allahverdi:2010xz}
R.~Allahverdi, R.~Brandenberger, F.-Y.~Cyr-Racine and A.~Mazumdar, \emph{{Reheating in Inflationary Cosmology: Theory and Applications}}, \href{https://doi.org/10.1146/annurev.nucl.012809.104511}{\emph{Ann. Rev. Nucl. Part. Sci.} {\bfseries 60} (2010) 27} [\href{https://arxiv.org/abs/1001.2600}{{\ttfamily 1001.2600}}].

\bibitem{Amin:2014eta}
M.A.~Amin, M.P.~Hertzberg, D.I.~Kaiser and J.~Karouby, \emph{{Nonperturbative Dynamics Of Reheating After Inflation: A Review}}, \href{https://doi.org/10.1142/S0218271815300037}{\emph{Int. J. Mod. Phys. D} {\bfseries 24} (2014) 1530003} [\href{https://arxiv.org/abs/1410.3808}{{\ttfamily 1410.3808}}].

\bibitem{Bezrukov:2007ep}
F.L.~Bezrukov and M.~Shaposhnikov, \emph{{The Standard Model Higgs boson as the inflaton}}, \href{https://doi.org/10.1016/j.physletb.2007.11.072}{\emph{Phys. Lett. B} {\bfseries 659} (2008) 703} [\href{https://arxiv.org/abs/0710.3755}{{\ttfamily 0710.3755}}].

\bibitem{Rubio:2018ogq}
J.~Rubio, \emph{{Higgs inflation}}, \href{https://doi.org/10.3389/fspas.2018.00050}{\emph{Front. Astron. Space Sci.} {\bfseries 5} (2019) 50} [\href{https://arxiv.org/abs/1807.02376}{{\ttfamily 1807.02376}}].

\bibitem{Barbon:2009ya}
J.L.F.~Barbon and J.R.~Espinosa, \emph{{On the Naturalness of Higgs Inflation}}, \href{https://doi.org/10.1103/PhysRevD.79.081302}{\emph{Phys. Rev. D} {\bfseries 79} (2009) 081302} [\href{https://arxiv.org/abs/0903.0355}{{\ttfamily 0903.0355}}].

\bibitem{Bezrukov:2009db}
F.~Bezrukov and M.~Shaposhnikov, \emph{{Standard Model Higgs boson mass from inflation: Two loop analysis}}, \href{https://doi.org/10.1088/1126-6708/2009/07/089}{\emph{JHEP} {\bfseries 07} (2009) 089} [\href{https://arxiv.org/abs/0904.1537}{{\ttfamily 0904.1537}}].

\bibitem{Burgess:2010zq}
C.P.~Burgess, H.M.~Lee and M.~Trott, \emph{{Comment on Higgs Inflation and Naturalness}}, \href{https://doi.org/10.1007/JHEP07(2010)007}{\emph{JHEP} {\bfseries 07} (2010) 007} [\href{https://arxiv.org/abs/1002.2730}{{\ttfamily 1002.2730}}].

\bibitem{Bezrukov:2010jz}
F.~Bezrukov, A.~Magnin, M.~Shaposhnikov and S.~Sibiryakov, \emph{{Higgs inflation: consistency and generalisations}}, \href{https://doi.org/10.1007/JHEP01(2011)016}{\emph{JHEP} {\bfseries 01} (2011) 016} [\href{https://arxiv.org/abs/1008.5157}{{\ttfamily 1008.5157}}].

\bibitem{Giudice:2010ka}
G.F.~Giudice and H.M.~Lee, \emph{{Unitarizing Higgs Inflation}}, \href{https://doi.org/10.1016/j.physletb.2010.10.035}{\emph{Phys. Lett. B} {\bfseries 694} (2011) 294} [\href{https://arxiv.org/abs/1010.1417}{{\ttfamily 1010.1417}}].

\bibitem{Bezrukov:2014bra}
F.~Bezrukov and M.~Shaposhnikov, \emph{{Higgs inflation at the critical point}}, \href{https://doi.org/10.1016/j.physletb.2014.05.074}{\emph{Phys. Lett. B} {\bfseries 734} (2014) 249} [\href{https://arxiv.org/abs/1403.6078}{{\ttfamily 1403.6078}}].

\bibitem{Hamada:2014iga}
Y.~Hamada, H.~Kawai, K.-y.~Oda and S.C.~Park, \emph{{Higgs Inflation is Still Alive after the Results from BICEP2}}, \href{https://doi.org/10.1103/PhysRevLett.112.241301}{\emph{Phys. Rev. Lett.} {\bfseries 112} (2014) 241301} [\href{https://arxiv.org/abs/1403.5043}{{\ttfamily 1403.5043}}].

\bibitem{Hamada:2014wna}
Y.~Hamada, H.~Kawai, K.-y.~Oda and S.C.~Park, \emph{{Higgs inflation from Standard Model criticality}}, \href{https://doi.org/10.1103/PhysRevD.91.053008}{\emph{Phys. Rev. D} {\bfseries 91} (2015) 053008} [\href{https://arxiv.org/abs/1408.4864}{{\ttfamily 1408.4864}}].

\bibitem{George:2015nza}
D.P.~George, S.~Mooij and M.~Postma, \emph{{Quantum corrections in Higgs inflation: the Standard Model case}}, \href{https://doi.org/10.1088/1475-7516/2016/04/006}{\emph{JCAP} {\bfseries 04} (2016) 006} [\href{https://arxiv.org/abs/1508.04660}{{\ttfamily 1508.04660}}].

\bibitem{Fumagalli:2016lls}
J.~Fumagalli and M.~Postma, \emph{{UV (in)sensitivity of Higgs inflation}}, \href{https://doi.org/10.1007/JHEP05(2016)049}{\emph{JHEP} {\bfseries 05} (2016) 049} [\href{https://arxiv.org/abs/1602.07234}{{\ttfamily 1602.07234}}].

\bibitem{Bezrukov:2017dyv}
F.~Bezrukov, M.~Pauly and J.~Rubio, \emph{{On the robustness of the primordial power spectrum in renormalized Higgs inflation}}, \href{https://doi.org/10.1088/1475-7516/2018/02/040}{\emph{JCAP} {\bfseries 02} (2018) 040} [\href{https://arxiv.org/abs/1706.05007}{{\ttfamily 1706.05007}}].

\bibitem{Bauer:2008zj}
F.~Bauer and D.A.~Demir, \emph{{Inflation with Non-Minimal Coupling: Metric versus Palatini Formulations}}, \href{https://doi.org/10.1016/j.physletb.2008.06.014}{\emph{Phys. Lett. B} {\bfseries 665} (2008) 222} [\href{https://arxiv.org/abs/0803.2664}{{\ttfamily 0803.2664}}].

\bibitem{Bauer:2010jg}
F.~Bauer and D.A.~Demir, \emph{{Higgs-Palatini Inflation and Unitarity}}, \href{https://doi.org/10.1016/j.physletb.2011.03.042}{\emph{Phys. Lett. B} {\bfseries 698} (2011) 425} [\href{https://arxiv.org/abs/1012.2900}{{\ttfamily 1012.2900}}].

\bibitem{Rasanen:2017ivk}
S.~Rasanen and P.~Wahlman, \emph{{Higgs inflation with loop corrections in the Palatini formulation}}, \href{https://doi.org/10.1088/1475-7516/2017/11/047}{\emph{JCAP} {\bfseries 11} (2017) 047} [\href{https://arxiv.org/abs/1709.07853}{{\ttfamily 1709.07853}}].

\bibitem{Enckell:2018kkc}
V.-M.~Enckell, K.~Enqvist, S.~Rasanen and E.~Tomberg, \emph{{Higgs inflation at the hilltop}}, \href{https://doi.org/10.1088/1475-7516/2018/06/005}{\emph{JCAP} {\bfseries 06} (2018) 005} [\href{https://arxiv.org/abs/1802.09299}{{\ttfamily 1802.09299}}].

\bibitem{Rasanen:2018fom}
S.~Rasanen and E.~Tomberg, \emph{{Planck scale black hole dark matter from Higgs inflation}}, \href{https://doi.org/10.1088/1475-7516/2019/01/038}{\emph{JCAP} {\bfseries 01} (2019) 038} [\href{https://arxiv.org/abs/1810.12608}{{\ttfamily 1810.12608}}].

\bibitem{Gialamas:2019nly}
I.D.~Gialamas and A.B.~Lahanas, \emph{{Reheating in $R^2$ Palatini inflationary models}}, \href{https://doi.org/10.1103/PhysRevD.101.084007}{\emph{Phys. Rev. D} {\bfseries 101} (2020) 084007} [\href{https://arxiv.org/abs/1911.11513}{{\ttfamily 1911.11513}}].

\bibitem{Shaposhnikov:2020fdv}
M.~Shaposhnikov, A.~Shkerin and S.~Zell, \emph{{Quantum Effects in Palatini Higgs Inflation}}, \href{https://doi.org/10.1088/1475-7516/2020/07/064}{\emph{JCAP} {\bfseries 07} (2020) 064} [\href{https://arxiv.org/abs/2002.07105}{{\ttfamily 2002.07105}}].

\bibitem{Annala:2021zdt}
J.~Annala and S.~Rasanen, \emph{{Inflation with R (\ensuremath{\alpha}\ensuremath{\beta}) terms in the Palatini formulation}}, \href{https://doi.org/10.1088/1475-7516/2021/09/032}{\emph{JCAP} {\bfseries 09} (2021) 032} [\href{https://arxiv.org/abs/2106.12422}{{\ttfamily 2106.12422}}].

\bibitem{Raatikainen:2019qey}
S.~Raatikainen and S.~Rasanen, \emph{{Higgs inflation and teleparallel gravity}}, \href{https://doi.org/10.1088/1475-7516/2019/12/021}{\emph{JCAP} {\bfseries 12} (2019) 021} [\href{https://arxiv.org/abs/1910.03488}{{\ttfamily 1910.03488}}].

\bibitem{Langvik:2020nrs}
M.~L\r{a}ngvik, J.-M.~Ojanper\"a, S.~Raatikainen and S.~R\"as\"anen, \emph{{Higgs inflation with the Holst and the Nieh\textendash{}Yan term}}, \href{https://doi.org/10.1103/PhysRevD.103.083514}{\emph{Phys. Rev. D} {\bfseries 103} (2021) 083514} [\href{https://arxiv.org/abs/2007.12595}{{\ttfamily 2007.12595}}].

\bibitem{Shaposhnikov:2020gts}
M.~Shaposhnikov, A.~Shkerin, I.~Timiryasov and S.~Zell, \emph{{Higgs inflation in Einstein-Cartan gravity}}, \href{https://doi.org/10.1088/1475-7516/2021/02/008}{\emph{JCAP} {\bfseries 02} (2021) 008} [\href{https://arxiv.org/abs/2007.14978}{{\ttfamily 2007.14978}}].

\bibitem{Shaposhnikov:2020aen}
M.~Shaposhnikov, A.~Shkerin, I.~Timiryasov and S.~Zell, \emph{{Einstein-Cartan Portal to Dark Matter}}, \href{https://doi.org/10.1103/PhysRevLett.127.169901}{\emph{Phys. Rev. Lett.} {\bfseries 126} (2021) 161301} [\href{https://arxiv.org/abs/2008.11686}{{\ttfamily 2008.11686}}].

\bibitem{Karananas:2021zkl}
G.K.~Karananas, M.~Shaposhnikov, A.~Shkerin and S.~Zell, \emph{{Matter matters in Einstein-Cartan gravity}}, \href{https://doi.org/10.1103/PhysRevD.104.064036}{\emph{Phys. Rev. D} {\bfseries 104} (2021) 064036} [\href{https://arxiv.org/abs/2106.13811}{{\ttfamily 2106.13811}}].

\bibitem{Azri:2017uor}
H.~Azri and D.~Demir, \emph{{Affine Inflation}}, \href{https://doi.org/10.1103/PhysRevD.95.124007}{\emph{Phys. Rev. D} {\bfseries 95} (2017) 124007} [\href{https://arxiv.org/abs/1705.05822}{{\ttfamily 1705.05822}}].

\bibitem{Rigouzzo:2022yan}
C.~Rigouzzo and S.~Zell, \emph{{Coupling metric-affine gravity to a Higgs-like scalar field}}, \href{https://doi.org/10.1103/PhysRevD.106.024015}{\emph{Phys. Rev. D} {\bfseries 106} (2022) 024015} [\href{https://arxiv.org/abs/2204.03003}{{\ttfamily 2204.03003}}].

\bibitem{Gialamas:2022xtt}
I.D.~Gialamas and K.~Tamvakis, \emph{{Inflation in metric-affine quadratic gravity}}, \href{https://doi.org/10.1088/1475-7516/2023/03/042}{\emph{JCAP} {\bfseries 03} (2023) 042} [\href{https://arxiv.org/abs/2212.09896}{{\ttfamily 2212.09896}}].

\bibitem{Shaposhnikov:2008xb}
M.~Shaposhnikov and D.~Zenhausern, \emph{{Scale invariance, unimodular gravity and dark energy}}, \href{https://doi.org/10.1016/j.physletb.2008.11.054}{\emph{Phys. Lett. B} {\bfseries 671} (2009) 187} [\href{https://arxiv.org/abs/0809.3395}{{\ttfamily 0809.3395}}].

\bibitem{Garcia-Bellido:2011kqb}
J.~Garcia-Bellido, J.~Rubio, M.~Shaposhnikov and D.~Zenhausern, \emph{{Higgs-Dilaton Cosmology: From the Early to the Late Universe}}, \href{https://doi.org/10.1103/PhysRevD.84.123504}{\emph{Phys. Rev. D} {\bfseries 84} (2011) 123504} [\href{https://arxiv.org/abs/1107.2163}{{\ttfamily 1107.2163}}].

\bibitem{Garcia-Bellido:2012npk}
J.~Garcia-Bellido, J.~Rubio and M.~Shaposhnikov, \emph{{Higgs-Dilaton cosmology: Are there extra relativistic species?}}, \href{https://doi.org/10.1016/j.physletb.2012.10.075}{\emph{Phys. Lett. B} {\bfseries 718} (2012) 507} [\href{https://arxiv.org/abs/1209.2119}{{\ttfamily 1209.2119}}].

\bibitem{Bezrukov:2012hx}
F.~Bezrukov, G.K.~Karananas, J.~Rubio and M.~Shaposhnikov, \emph{{Higgs-Dilaton Cosmology: an effective field theory approach}}, \href{https://doi.org/10.1103/PhysRevD.87.096001}{\emph{Phys. Rev. D} {\bfseries 87} (2013) 096001} [\href{https://arxiv.org/abs/1212.4148}{{\ttfamily 1212.4148}}].

\bibitem{Rubio:2014wta}
J.~Rubio and M.~Shaposhnikov, \emph{{Higgs-Dilaton cosmology: Universality versus criticality}}, \href{https://doi.org/10.1103/PhysRevD.90.027307}{\emph{Phys. Rev. D} {\bfseries 90} (2014) 027307} [\href{https://arxiv.org/abs/1406.5182}{{\ttfamily 1406.5182}}].

\bibitem{Trashorras:2016azl}
M.~Trashorras, S.~Nesseris and J.~Garcia-Bellido, \emph{{Cosmological Constraints on Higgs-Dilaton Inflation}}, \href{https://doi.org/10.1103/PhysRevD.94.063511}{\emph{Phys. Rev. D} {\bfseries 94} (2016) 063511} [\href{https://arxiv.org/abs/1604.06760}{{\ttfamily 1604.06760}}].

\bibitem{Casas:2017wjh}
S.~Casas, M.~Pauly and J.~Rubio, \emph{{Higgs-dilaton cosmology: An inflation\textendash{}dark-energy connection and forecasts for future galaxy surveys}}, \href{https://doi.org/10.1103/PhysRevD.97.043520}{\emph{Phys. Rev. D} {\bfseries 97} (2018) 043520} [\href{https://arxiv.org/abs/1712.04956}{{\ttfamily 1712.04956}}].

\bibitem{Tokareva:2017nng}
A.~Tokareva, \emph{{A minimal scale invariant axion solution to the strong CP-problem}}, \href{https://doi.org/10.1140/epjc/s10052-018-5883-0}{\emph{Eur. Phys. J. C} {\bfseries 78} (2018) 423} [\href{https://arxiv.org/abs/1705.10836}{{\ttfamily 1705.10836}}].

\bibitem{Casas:2018fum}
S.~Casas, G.K.~Karananas, M.~Pauly and J.~Rubio, \emph{{Scale-invariant alternatives to general relativity. III. The inflation-dark energy connection}}, \href{https://doi.org/10.1103/PhysRevD.99.063512}{\emph{Phys. Rev. D} {\bfseries 99} (2019) 063512} [\href{https://arxiv.org/abs/1811.05984}{{\ttfamily 1811.05984}}].

\bibitem{Shaposhnikov:2018jag}
M.~Shaposhnikov and A.~Shkerin, \emph{{Gravity, Scale Invariance and the Hierarchy Problem}}, \href{https://doi.org/10.1007/JHEP10(2018)024}{\emph{JHEP} {\bfseries 10} (2018) 024} [\href{https://arxiv.org/abs/1804.06376}{{\ttfamily 1804.06376}}].

\bibitem{Herrero-Valea:2019hde}
M.~Herrero-Valea, I.~Timiryasov and A.~Tokareva, \emph{{To Positivity and Beyond, where Higgs-Dilaton Inflation has never gone before}}, \href{https://doi.org/10.1088/1475-7516/2019/11/042}{\emph{JCAP} {\bfseries 11} (2019) 042} [\href{https://arxiv.org/abs/1905.08816}{{\ttfamily 1905.08816}}].

\bibitem{Karananas:2020qkp}
G.K.~Karananas, M.~Michel and J.~Rubio, \emph{{One residue to rule them all: Electroweak symmetry breaking, inflation and field-space geometry}}, \href{https://doi.org/10.1016/j.physletb.2020.135876}{\emph{Phys. Lett. B} {\bfseries 811} (2020) 135876} [\href{https://arxiv.org/abs/2006.11290}{{\ttfamily 2006.11290}}].

\bibitem{Rubio:2020zht}
J.~Rubio, \emph{{Scale symmetry, the Higgs and the Cosmos}}, \href{https://doi.org/10.22323/1.376.0074}{\emph{PoS} {\bfseries CORFU2019} (2020) 074} [\href{https://arxiv.org/abs/2004.00039}{{\ttfamily 2004.00039}}].

\bibitem{Shaposhnikov:2020frq}
M.~Shaposhnikov, A.~Shkerin, I.~Timiryasov and S.~Zell, \emph{{Einstein-Cartan gravity, matter, and scale-invariant generalization~}}, \href{https://doi.org/10.1007/JHEP08(2021)162}{\emph{JHEP} {\bfseries 10} (2020) 177} [\href{https://arxiv.org/abs/2007.16158}{{\ttfamily 2007.16158}}].

\bibitem{Karananas:2021gco}
G.K.~Karananas, M.~Shaposhnikov, A.~Shkerin and S.~Zell, \emph{{Scale and Weyl Invariance in Einstein-Cartan Gravity}},  \href{https://arxiv.org/abs/2108.05897}{{\ttfamily 2108.05897}}.

\bibitem{Gialamas:2021enw}
I.D.~Gialamas, A.~Karam, T.D.~Pappas and V.C.~Spanos, \emph{{Scale-invariant quadratic gravity and inflation in the Palatini formalism}}, \href{https://doi.org/10.1103/PhysRevD.104.023521}{\emph{Phys. Rev. D} {\bfseries 104} (2021) 023521} [\href{https://arxiv.org/abs/2104.04550}{{\ttfamily 2104.04550}}].

\bibitem{Piani:2022gon}
M.~Piani and J.~Rubio, \emph{{Higgs-Dilaton inflation in Einstein-Cartan gravity}}, \href{https://doi.org/10.1088/1475-7516/2022/05/009}{\emph{JCAP} {\bfseries 05} (2022) 009} [\href{https://arxiv.org/abs/2202.04665}{{\ttfamily 2202.04665}}].

\bibitem{Belokon:2022pqf}
A.I.~Belokon and A.~Tokareva, \emph{{Higgs-dilaton model revisited: can dilaton act as QCD axion?}},  \href{https://arxiv.org/abs/2212.06739}{{\ttfamily 2212.06739}}.

\bibitem{Bezrukov:2008ut}
F.~Bezrukov, D.~Gorbunov and M.~Shaposhnikov, \emph{{On initial conditions for the Hot Big Bang}}, \href{https://doi.org/10.1088/1475-7516/2009/06/029}{\emph{JCAP} {\bfseries 06} (2009) 029} [\href{https://arxiv.org/abs/0812.3622}{{\ttfamily 0812.3622}}].

\bibitem{Garcia-Bellido:2008ycs}
J.~Garcia-Bellido, D.G.~Figueroa and J.~Rubio, \emph{{Preheating in the Standard Model with the Higgs-Inflaton coupled to gravity}}, \href{https://doi.org/10.1103/PhysRevD.79.063531}{\emph{Phys. Rev. D} {\bfseries 79} (2009) 063531} [\href{https://arxiv.org/abs/0812.4624}{{\ttfamily 0812.4624}}].

\bibitem{Bezrukov:2014ipa}
F.~Bezrukov, J.~Rubio and M.~Shaposhnikov, \emph{{Living beyond the edge: Higgs inflation and vacuum metastability}}, \href{https://doi.org/10.1103/PhysRevD.92.083512}{\emph{Phys. Rev. D} {\bfseries 92} (2015) 083512} [\href{https://arxiv.org/abs/1412.3811}{{\ttfamily 1412.3811}}].

\bibitem{Repond:2016sol}
J.~Repond and J.~Rubio, \emph{{Combined Preheating on the lattice with applications to Higgs inflation}}, \href{https://doi.org/10.1088/1475-7516/2016/07/043}{\emph{JCAP} {\bfseries 07} (2016) 043} [\href{https://arxiv.org/abs/1604.08238}{{\ttfamily 1604.08238}}].

\bibitem{DeCross:2015uza}
M.P.~DeCross, D.I.~Kaiser, A.~Prabhu, C.~Prescod-Weinstein and E.I.~Sfakianakis, \emph{{Preheating after Multifield Inflation with Nonminimal Couplings, I: Covariant Formalism and Attractor Behavior}}, \href{https://doi.org/10.1103/PhysRevD.97.023526}{\emph{Phys. Rev. D} {\bfseries 97} (2018) 023526} [\href{https://arxiv.org/abs/1510.08553}{{\ttfamily 1510.08553}}].

\bibitem{DeCross:2016fdz}
M.P.~DeCross, D.I.~Kaiser, A.~Prabhu, C.~Prescod-Weinstein and E.I.~Sfakianakis, \emph{{Preheating after multifield inflation with nonminimal couplings, II: Resonance Structure}}, \href{https://doi.org/10.1103/PhysRevD.97.023527}{\emph{Phys. Rev. D} {\bfseries 97} (2018) 023527} [\href{https://arxiv.org/abs/1610.08868}{{\ttfamily 1610.08868}}].

\bibitem{DeCross:2016cbs}
M.P.~DeCross, D.I.~Kaiser, A.~Prabhu, C.~Prescod-Weinstein and E.I.~Sfakianakis, \emph{{Preheating after multifield inflation with nonminimal couplings, III: Dynamical spacetime results}}, \href{https://doi.org/10.1103/PhysRevD.97.023528}{\emph{Phys. Rev. D} {\bfseries 97} (2018) 023528} [\href{https://arxiv.org/abs/1610.08916}{{\ttfamily 1610.08916}}].

\bibitem{Ema:2016dny}
Y.~Ema, R.~Jinno, K.~Mukaida and K.~Nakayama, \emph{{Violent Preheating in Inflation with Nonminimal Coupling}}, \href{https://doi.org/10.1088/1475-7516/2017/02/045}{\emph{JCAP} {\bfseries 02} (2017) 045} [\href{https://arxiv.org/abs/1609.05209}{{\ttfamily 1609.05209}}].

\bibitem{Sfakianakis:2018lzf}
E.I.~Sfakianakis and J.~van~de Vis, \emph{{Preheating after Higgs Inflation: Self-Resonance and Gauge boson production}}, \href{https://doi.org/10.1103/PhysRevD.99.083519}{\emph{Phys. Rev. D} {\bfseries 99} (2019) 083519} [\href{https://arxiv.org/abs/1810.01304}{{\ttfamily 1810.01304}}].

\bibitem{Rubio:2019ypq}
J.~Rubio and E.S.~Tomberg, \emph{{Preheating in Palatini Higgs inflation}}, \href{https://doi.org/10.1088/1475-7516/2019/04/021}{\emph{JCAP} {\bfseries 04} (2019) 021} [\href{https://arxiv.org/abs/1902.10148}{{\ttfamily 1902.10148}}].

\bibitem{Dux:2022kuk}
F.~Dux, A.~Florio, J.~Klari\'c, A.~Shkerin and I.~Timiryasov, \emph{{Preheating in Palatini Higgs inflation on the lattice}},  \href{https://arxiv.org/abs/2203.13286}{{\ttfamily 2203.13286}}.

\bibitem{Rigouzzo:2023sbb}
C.~Rigouzzo and S.~Zell, \emph{{Coupling Metric-Affine Gravity to the Standard Model and Dark Matter Fermions}},  \href{https://arxiv.org/abs/2306.13134}{{\ttfamily 2306.13134}}.

\bibitem{Utiyama:1956sy}
R.~Utiyama, \emph{{Invariant theoretical interpretation of interaction}}, \href{https://doi.org/10.1103/PhysRev.101.1597}{\emph{Phys. Rev.} {\bfseries 101} (1956) 1597}.

\bibitem{Kibble:1961ba}
T.W.B.~Kibble, \emph{{Lorentz invariance and the gravitational field}}, \href{https://doi.org/10.1063/1.1703702}{\emph{J. Math. Phys.} {\bfseries 2} (1961) 212}.

\bibitem{Hehl:1976kj}
F.W.~Hehl, P.~Von Der~Heyde, G.D.~Kerlick and J.M.~Nester, \emph{{General Relativity with Spin and Torsion: Foundations and Prospects}}, \href{https://doi.org/10.1103/RevModPhys.48.393}{\emph{Rev. Mod. Phys.} {\bfseries 48} (1976) 393}.

\bibitem{Shapiro:2001rz}
I.L.~Shapiro, \emph{{Physical aspects of the space-time torsion}}, \href{https://doi.org/10.1016/S0370-1573(01)00030-8}{\emph{Phys. Rept.} {\bfseries 357} (2002) 113} [\href{https://arxiv.org/abs/hep-th/0103093}{{\ttfamily hep-th/0103093}}].

\bibitem{Ferraris1982}
M.~Ferraris, M.~Francaviglia and C.~Reina, \emph{{Variational formulation of general relativity from 1915 to 1925 Palatini's method discovered by Einstein in 1925}}, {\emph{General Relativity and Gravitation} {\bfseries 14} (1982) 243}.

\bibitem{Nieh:1981ww}
H.T.~Nieh and M.L.~Yan, \emph{{An Identity in Riemann-cartan Geometry}}, \href{https://doi.org/10.1063/1.525379}{\emph{J. Math. Phys.} {\bfseries 23} (1982) 373}.

\bibitem{Nieh:2008btw}
H.T.~Nieh, \emph{{A Torsional Topological Invariant}},  in \emph{{Conference in Honor of C.N. Yang's 85th Birthday}: {Statistical Physics, High Energy, Condensed Matter and Mathematical Physics}}, pp.~29--37, 2008, \href{https://doi.org/10.1142/9789812794185_0003}{DOI} [\href{https://arxiv.org/abs/1309.0915}{{\ttfamily 1309.0915}}].

\bibitem{Kallosh:2021mnu}
R.~Kallosh and A.~Linde, \emph{{BICEP/Keck and cosmological attractors}}, \href{https://doi.org/10.1088/1475-7516/2021/12/008}{\emph{JCAP} {\bfseries 12} (2021) 008} [\href{https://arxiv.org/abs/2110.10902}{{\ttfamily 2110.10902}}].

\bibitem{BICEPKeck:2022mhb}
{\scshape BICEP/Keck} collaboration, \emph{{The Latest Constraints on Inflationary B-modes from the BICEP/Keck Telescopes}},  in \emph{{56th Rencontres de Moriond on Cosmology}}, 3, 2022 [\href{https://arxiv.org/abs/2203.16556}{{\ttfamily 2203.16556}}].

\bibitem{Freidel:2005sn}
L.~Freidel, D.~Minic and T.~Takeuchi, \emph{{Quantum gravity, torsion, parity violation and all that}}, \href{https://doi.org/10.1103/PhysRevD.72.104002}{\emph{Phys. Rev. D} {\bfseries 72} (2005) 104002} [\href{https://arxiv.org/abs/hep-th/0507253}{{\ttfamily hep-th/0507253}}].

\bibitem{Alexandrov:2008iy}
S.~Alexandrov, \emph{{Immirzi parameter and fermions with non-minimal coupling}}, \href{https://doi.org/10.1088/0264-9381/25/14/145012}{\emph{Class. Quant. Grav.} {\bfseries 25} (2008) 145012} [\href{https://arxiv.org/abs/0802.1221}{{\ttfamily 0802.1221}}].

\bibitem{Diakonov:2011fs}
D.~Diakonov, A.G.~Tumanov and A.A.~Vladimirov, \emph{{Low-energy General Relativity with torsion: A Systematic derivative expansion}}, \href{https://doi.org/10.1103/PhysRevD.84.124042}{\emph{Phys. Rev. D} {\bfseries 84} (2011) 124042} [\href{https://arxiv.org/abs/1104.2432}{{\ttfamily 1104.2432}}].

\bibitem{Magueijo:2012ug}
J.a.~Magueijo, T.G.~Zlosnik and T.W.B.~Kibble, \emph{{Cosmology with a spin}}, \href{https://doi.org/10.1103/PhysRevD.87.063504}{\emph{Phys. Rev. D} {\bfseries 87} (2013) 063504} [\href{https://arxiv.org/abs/1212.0585}{{\ttfamily 1212.0585}}].

\bibitem{ATLAS:2019guf}
{\scshape ATLAS} collaboration, \emph{{Measurement of the top-quark mass in $t\bar{t}+1$-jet events collected with the ATLAS detector in $pp$ collisions at $\sqrt{s}=8$ TeV}}, \href{https://doi.org/10.1007/JHEP11(2019)150}{\emph{JHEP} {\bfseries 11} (2019) 150} [\href{https://arxiv.org/abs/1905.02302}{{\ttfamily 1905.02302}}].

\bibitem{CMS:2019esx}
{\scshape CMS} collaboration, \emph{{Measurement of $\mathrm{t\bar t}$ normalised multi-differential cross sections in pp collisions at $\sqrt s=13$ TeV, and simultaneous determination of the strong coupling strength, top quark pole mass, and parton distribution functions}}, \href{https://doi.org/10.1140/epjc/s10052-020-7917-7}{\emph{Eur. Phys. J. C} {\bfseries 80} (2020) 658} [\href{https://arxiv.org/abs/1904.05237}{{\ttfamily 1904.05237}}].

\bibitem{Rubio:2015zia}
J.~Rubio, \emph{{Higgs inflation and vacuum stability}}, \href{https://doi.org/10.1088/1742-6596/631/1/012032}{\emph{J. Phys. Conf. Ser.} {\bfseries 631} (2015) 012032} [\href{https://arxiv.org/abs/1502.07952}{{\ttfamily 1502.07952}}].

\bibitem{Greene:2000ew}
P.B.~Greene and L.~Kofman, \emph{{On the theory of fermionic preheating}}, \href{https://doi.org/10.1103/PhysRevD.62.123516}{\emph{Phys. Rev. D} {\bfseries 62} (2000) 123516} [\href{https://arxiv.org/abs/hep-ph/0003018}{{\ttfamily hep-ph/0003018}}].

\bibitem{Peloso:2000hy}
M.~Peloso and L.~Sorbo, \emph{{Preheating of massive fermions after inflation: Analytical results}}, \href{https://doi.org/10.1088/1126-6708/2000/05/016}{\emph{JHEP} {\bfseries 05} (2000) 016} [\href{https://arxiv.org/abs/hep-ph/0003045}{{\ttfamily hep-ph/0003045}}].

\bibitem{Bogolyubsky:1976yu}
I.L.~Bogolyubsky and V.G.~Makhankov, \emph{{Lifetime of Pulsating Solitons in Some Classical Models}}, {\emph{Pisma Zh. Eksp. Teor. Fiz.} {\bfseries 24} (1976) 15}.

\bibitem{Gleiser:1993pt}
M.~Gleiser, \emph{{Pseudostable bubbles}}, \href{https://doi.org/10.1103/PhysRevD.49.2978}{\emph{Phys. Rev. D} {\bfseries 49} (1994) 2978} [\href{https://arxiv.org/abs/hep-ph/9308279}{{\ttfamily hep-ph/9308279}}].

\bibitem{Copeland:1995fq}
E.J.~Copeland, M.~Gleiser and H.R.~Muller, \emph{{Oscillons: Resonant configurations during bubble collapse}}, \href{https://doi.org/10.1103/PhysRevD.52.1920}{\emph{Phys. Rev. D} {\bfseries 52} (1995) 1920} [\href{https://arxiv.org/abs/hep-ph/9503217}{{\ttfamily hep-ph/9503217}}].

\bibitem{Amin:2010dc}
M.A.~Amin, R.~Easther and H.~Finkel, \emph{{Inflaton Fragmentation and Oscillon Formation in Three Dimensions}}, \href{https://doi.org/10.1088/1475-7516/2010/12/001}{\emph{JCAP} {\bfseries 12} (2010) 001} [\href{https://arxiv.org/abs/1009.2505}{{\ttfamily 1009.2505}}].

\bibitem{Amin:2011hj}
M.A.~Amin, R.~Easther, H.~Finkel, R.~Flauger and M.P.~Hertzberg, \emph{{Oscillons After Inflation}}, \href{https://doi.org/10.1103/PhysRevLett.108.241302}{\emph{Phys. Rev. Lett.} {\bfseries 108} (2012) 241302} [\href{https://arxiv.org/abs/1106.3335}{{\ttfamily 1106.3335}}].

\bibitem{Lozanov:2017hjm}
K.D.~Lozanov and M.A.~Amin, \emph{{Self-resonance after inflation: oscillons, transients and radiation domination}}, \href{https://doi.org/10.1103/PhysRevD.97.023533}{\emph{Phys. Rev. D} {\bfseries 97} (2018) 023533} [\href{https://arxiv.org/abs/1710.06851}{{\ttfamily 1710.06851}}].

\bibitem{Lozanov:2019ylm}
K.D.~Lozanov and M.A.~Amin, \emph{{Gravitational perturbations from oscillons and transients after inflation}}, \href{https://doi.org/10.1103/PhysRevD.99.123504}{\emph{Phys. Rev. D} {\bfseries 99} (2019) 123504} [\href{https://arxiv.org/abs/1902.06736}{{\ttfamily 1902.06736}}].

\bibitem{Hasegawa:2017iay}
F.~Hasegawa and J.-P.~Hong, \emph{{Inflaton fragmentation in E-models of cosmological $\alpha$-attractors}}, \href{https://doi.org/10.1103/PhysRevD.97.083514}{\emph{Phys. Rev. D} {\bfseries 97} (2018) 083514} [\href{https://arxiv.org/abs/1710.07487}{{\ttfamily 1710.07487}}].

\bibitem{Mahbub:2023faw}
R.~Mahbub and S.S.~Mishra, \emph{{Oscillon formation from preheating in asymmetric inflationary potentials}},  \href{https://arxiv.org/abs/2303.07503}{{\ttfamily 2303.07503}}.

\bibitem{Sang:2019ndv}
Y.~Sang and Q.-G.~Huang, \emph{{Stochastic Gravitational-Wave Background from Axion-Monodromy Oscillons in String Theory During Preheating}}, \href{https://doi.org/10.1103/PhysRevD.100.063516}{\emph{Phys. Rev. D} {\bfseries 100} (2019) 063516} [\href{https://arxiv.org/abs/1905.00371}{{\ttfamily 1905.00371}}].

\bibitem{Sang:2020kpd}
Y.~Sang and Q.-G.~Huang, \emph{{Oscillons during Dirac-Born-Infeld preheating}}, \href{https://doi.org/10.1016/j.physletb.2021.136781}{\emph{Phys. Lett. B} {\bfseries 823} (2021) 136781} [\href{https://arxiv.org/abs/2012.14697}{{\ttfamily 2012.14697}}].

\bibitem{Figueroa:2020rrl}
D.G.~Figueroa, A.~Florio, F.~Torrenti and W.~Valkenburg, \emph{{The art of simulating the early Universe -- Part I}}, \href{https://doi.org/10.1088/1475-7516/2021/04/035}{\emph{JCAP} {\bfseries 04} (2021) 035} [\href{https://arxiv.org/abs/2006.15122}{{\ttfamily 2006.15122}}].

\bibitem{Figueroa:2021yhd}
D.G.~Figueroa, A.~Florio, F.~Torrenti and W.~Valkenburg, \emph{{CosmoLattice}},  \href{https://arxiv.org/abs/2102.01031}{{\ttfamily 2102.01031}}.

\bibitem{Hazumi:2019lys}
M.~Hazumi et~al., \emph{{LiteBIRD: A Satellite for the Studies of B-Mode Polarization and Inflation from Cosmic Background Radiation Detection}}, \href{https://doi.org/10.1007/s10909-019-02150-5}{\emph{J. Low Temp. Phys.} {\bfseries 194} (2019) 443}.

\bibitem{Allahverdi:2020bys}
R.~Allahverdi et~al., \emph{{The First Three Seconds: a Review of Possible Expansion Histories of the Early Universe}},  \href{https://arxiv.org/abs/2006.16182}{{\ttfamily 2006.16182}}.

\bibitem{tech}
J.~Baeza-Ballesteros, A.~Figueroa, Daniel G.and~Florio and N.~Loayza, ``{CosmoLattice Technical Note II: Gravitational waves}.'' \url{https://cosmolattice.net/assets/technical_notes/CosmoLattice_TechnicalNote_GWs.pdf}, 2022.

\bibitem{Kurkela:2011ti}
A.~Kurkela and G.D.~Moore, \emph{{Thermalization in Weakly Coupled Nonabelian Plasmas}}, \href{https://doi.org/10.1007/JHEP12(2011)044}{\emph{JHEP} {\bfseries 12} (2011) 044} [\href{https://arxiv.org/abs/1107.5050}{{\ttfamily 1107.5050}}].

\bibitem{Micha:2002ey}
R.~Micha and I.I.~Tkachev, \emph{{Relativistic turbulence: A Long way from preheating to equilibrium}}, \href{https://doi.org/10.1103/PhysRevLett.90.121301}{\emph{Phys. Rev. Lett.} {\bfseries 90} (2003) 121301} [\href{https://arxiv.org/abs/hep-ph/0210202}{{\ttfamily hep-ph/0210202}}].

\bibitem{Micha:2004bv}
R.~Micha and I.I.~Tkachev, \emph{{Turbulent thermalization}}, \href{https://doi.org/10.1103/PhysRevD.70.043538}{\emph{Phys. Rev. D} {\bfseries 70} (2004) 043538} [\href{https://arxiv.org/abs/hep-ph/0403101}{{\ttfamily hep-ph/0403101}}].

\bibitem{Dufaux:2007pt}
J.F.~Dufaux, A.~Bergman, G.N.~Felder, L.~Kofman and J.-P.~Uzan, \emph{{Theory and Numerics of Gravitational Waves from Preheating after Inflation}}, \href{https://doi.org/10.1103/PhysRevD.76.123517}{\emph{Phys. Rev. D} {\bfseries 76} (2007) 123517} [\href{https://arxiv.org/abs/0707.0875}{{\ttfamily 0707.0875}}].

\bibitem{Aggarwal:2020olq}
N.~Aggarwal et~al., \emph{{Challenges and opportunities of gravitational-wave searches at MHz to GHz frequencies}}, \href{https://doi.org/10.1007/s41114-021-00032-5}{\emph{Living Rev. Rel.} {\bfseries 24} (2021) 4} [\href{https://arxiv.org/abs/2011.12414}{{\ttfamily 2011.12414}}].

\bibitem{Figueroa:2015rqa}
D.G.~Figueroa, J.~Garcia-Bellido and F.~Torrenti, \emph{{Decay of the standard model Higgs field after inflation}}, \href{https://doi.org/10.1103/PhysRevD.92.083511}{\emph{Phys. Rev. D} {\bfseries 92} (2015) 083511} [\href{https://arxiv.org/abs/1504.04600}{{\ttfamily 1504.04600}}].

\bibitem{Enqvist:2014tta}
K.~Enqvist, S.~Nurmi and S.~Rusak, \emph{{Non-Abelian dynamics in the resonant decay of the Higgs after inflation}}, \href{https://doi.org/10.1088/1475-7516/2014/10/064}{\emph{JCAP} {\bfseries 10} (2014) 064} [\href{https://arxiv.org/abs/1404.3631}{{\ttfamily 1404.3631}}].

\bibitem{Enqvist:2015sua}
K.~Enqvist, S.~Nurmi, S.~Rusak and D.~Weir, \emph{{Lattice Calculation of the Decay of Primordial Higgs Condensate}}, \href{https://doi.org/10.1088/1475-7516/2016/02/057}{\emph{JCAP} {\bfseries 02} (2016) 057} [\href{https://arxiv.org/abs/1506.06895}{{\ttfamily 1506.06895}}].

\bibitem{Kofman:1997yn}
L.~Kofman, A.D.~Linde and A.A.~Starobinsky, \emph{{Towards the theory of reheating after inflation}}, \href{https://doi.org/10.1103/PhysRevD.56.3258}{\emph{Phys. Rev. D} {\bfseries 56} (1997) 3258} [\href{https://arxiv.org/abs/hep-ph/9704452}{{\ttfamily hep-ph/9704452}}].

\bibitem{Felder:1998vq}
G.N.~Felder, L.~Kofman and A.D.~Linde, \emph{{Instant preheating}}, \href{https://doi.org/10.1103/PhysRevD.59.123523}{\emph{Phys. Rev. D} {\bfseries 59} (1999) 123523} [\href{https://arxiv.org/abs/hep-ph/9812289}{{\ttfamily hep-ph/9812289}}].

\bibitem{Nielsen:1981hk}
H.B.~Nielsen and M.~Ninomiya, \emph{{No Go Theorem for Regularizing Chiral Fermions}}, \href{https://doi.org/10.1016/0370-2693(81)91026-1}{\emph{Phys. Lett. B} {\bfseries 105} (1981) 219}.

\bibitem{Fan:2021otj}
J.~Fan, K.D.~Lozanov and Q.~Lu, \emph{{Spillway Preheating}}, \href{https://doi.org/10.1007/JHEP05(2021)069}{\emph{JHEP} {\bfseries 05} (2021) 069} [\href{https://arxiv.org/abs/2101.11008}{{\ttfamily 2101.11008}}].

\bibitem{Blas:2011ac}
D.~Blas, M.~Shaposhnikov and D.~Zenhausern, \emph{{Scale-invariant alternatives to general relativity}}, \href{https://doi.org/10.1103/PhysRevD.84.044001}{\emph{Phys. Rev. D} {\bfseries 84} (2011) 044001} [\href{https://arxiv.org/abs/1104.1392}{{\ttfamily 1104.1392}}].

\bibitem{Segur:1987mg}
H.~Segur and M.D.~Kruskal, \emph{{Nonexistence of Small Amplitude Breather Solutions in $\phi^4$ Theory}}, \href{https://doi.org/10.1103/PhysRevLett.58.747}{\emph{Phys. Rev. Lett.} {\bfseries 58} (1987) 747}.

\bibitem{Gleiser:2009ys}
M.~Gleiser and D.~Sicilia, \emph{{A General Theory of Oscillon Dynamics}}, \href{https://doi.org/10.1103/PhysRevD.80.125037}{\emph{Phys. Rev. D} {\bfseries 80} (2009) 125037} [\href{https://arxiv.org/abs/0910.5922}{{\ttfamily 0910.5922}}].

\bibitem{Hertzberg:2010yz}
M.P.~Hertzberg, \emph{{Quantum Radiation of Oscillons}}, \href{https://doi.org/10.1103/PhysRevD.82.045022}{\emph{Phys. Rev. D} {\bfseries 82} (2010) 045022} [\href{https://arxiv.org/abs/1003.3459}{{\ttfamily 1003.3459}}].

\bibitem{Salmi:2012ta}
P.~Salmi and M.~Hindmarsh, \emph{{Radiation and Relaxation of Oscillons}}, \href{https://doi.org/10.1103/PhysRevD.85.085033}{\emph{Phys. Rev. D} {\bfseries 85} (2012) 085033} [\href{https://arxiv.org/abs/1201.1934}{{\ttfamily 1201.1934}}].

\bibitem{Evslin:2023qbv}
J.~Evslin, T.~Roma\'nczukiewicz and A.~Wereszczy\'nski, \emph{{Quantum oscillons may be long-lived}}, \href{https://doi.org/10.1007/JHEP08(2023)182}{\emph{JHEP} {\bfseries 08} (2023) 182} [\href{https://arxiv.org/abs/2305.18056}{{\ttfamily 2305.18056}}].

\bibitem{Cotner:2018vug}
E.~Cotner, A.~Kusenko and V.~Takhistov, \emph{{Primordial Black Holes from Inflaton Fragmentation into Oscillons}}, \href{https://doi.org/10.1103/PhysRevD.98.083513}{\emph{Phys. Rev. D} {\bfseries 98} (2018) 083513} [\href{https://arxiv.org/abs/1801.03321}{{\ttfamily 1801.03321}}].

\bibitem{Amin:2019ums}
M.A.~Amin and P.~Mocz, \emph{{Formation, gravitational clustering, and interactions of nonrelativistic solitons in an expanding universe}}, \href{https://doi.org/10.1103/PhysRevD.100.063507}{\emph{Phys. Rev. D} {\bfseries 100} (2019) 063507} [\href{https://arxiv.org/abs/1902.07261}{{\ttfamily 1902.07261}}].

\end{thebibliography}\endgroup
%%%%%%%%%%%%%%%%%%%%%%%%%%%%%%%%%%%%%%%%%%%%%%%%%%%%%%%%%%%%%%%%%%%%%%%%%%%%%%%%
\end{document}